\pdfoutput=1

\documentclass[twocolumn]{revtex4-2}

\usepackage{graphicx}
\usepackage{amsmath,amssymb}
\usepackage{color}
\usepackage{hyperref}

\begin{document}

\title{Reexamination of eccentricity scaling of elliptic flow and incomplete thermalization scenario in heavy-ion collisions at energies available at CERN Large Hadron Collider}

\author{Arun Kumar Yadav}
\email{ak.yadav@vecc.gov.in}
\affiliation{Variable Energy Cyclotron Centre, HBNI, 1/AF Bidhan Nagar, Kolkata 700 064, India}
\affiliation{Homi Bhabha National Institute, Anushakti Nagar, Mumbai 400094, India}

\author{Partha Pratim Bhaduri}
\email{partha.bhaduri@vecc.gov.in}
\affiliation{Variable Energy Cyclotron Centre, HBNI, 1/AF Bidhan Nagar, Kolkata 700 064, India}
\affiliation{Homi Bhabha National Institute, Anushakti Nagar, Mumbai 400094, India}

\author{Subhasis Chattopadhyay}
\affiliation{Variable Energy Cyclotron Centre, HBNI, 1/AF Bidhan Nagar, Kolkata 700 064, India}
\affiliation{Homi Bhabha National Institute, Anushakti Nagar, Mumbai 400094, India}

\date{\today}

\begin{abstract}
 In this article, we reexamine the formulation for extraction of Knudsen number ($K$), the ratio of shear viscosity to entropy density ($\eta/s$), within the incomplete thermalization scenario, using eccentricity scaling of elliptic flow ($v_{2}$) of final state hadrons. Data on centrality dependence of charged hadron $v_{2}$ in Xe-Xe and Pb-Pb collisions, measured by ALICE and CMS collaborations at LHC are analyzed for this purpose. Results have been compared using two different models of collision namely Glauber Monte Carlo and TRENTO. The measured $v_{2}$, even for $\sqrt{s_{NN}}=5.02$ TeV most central Pb-Pb collisions is found to be below the ideal hydrodynamic limit by at least $15 \%$. Extracted $\eta/s$ of the medium exhibit heavy dependence on the employed initial conditions. Impact of the Pade approximants based alternative formulation of the functional relation between $v_{2}$ and $K$ are also investigated in detail. Our studies reconfirm the importance of clarifying the initial state configuration as well as the inbuilt ambiguities in employing the transport approach based on incomplete equilibrium in a strongly coupled regime at LHC.
\keywords{elliptic flow \and thermalization \and LHC \and shear viscosity to entropy density ratio}
\end{abstract}
\maketitle
\section{Introduction}
\label{intro}
Ultra-relativistic heavy-ion collision experiments at the BNL Relativistic Heavy Ion Collider (RHIC) and at the CERN Large Hadron Collider (LHC) have performed confirmatory tests of producing and characterizing a strongly coupled quark-gluon plasma (sQGP) in the laboratory~\cite{BRAHMS:2004adc,PHOBOS:2004zne,STAR:2005gfr,PHENIX:2004vcz}. A major goal of the ongoing research activity is the precise quantification of the fundamental properties of this hot and dense fluid medium. Typically this is achieved by matching the experimental measurements to the computational models which mimic the full space-time evolution of the heavy-ion collisions. One of the important features of the plasma medium is the collective hydrodynamic expansion probed via the measurement of anisotropic flow of the produced hadrons. Quantified with coefficients $v_n$ it measures the momentum anisotropy of the final-state particles and shows sensitivity both to the initial geometry of the overlap zone and to the transport properties and Equation of State (EoS) of the produced matter. In non-central collisions, the overlap region typically has an approximate ellipsoidal shape, making elliptic flow, traditionally expressed as $v_2 = < cos2(\phi - \psi_{EP})>$, the dominant flow coefficient\cite{Voloshin:1994mz}. It develops due to the multiple re-scatterings among the constituents of the overlap zone between the colliding nuclei and the resulting pressure gradient that converts the initial state spatial anisotropy to final state momentum anisotropy. 
Observation of large $v_2$ is one of the highlights of the heavy-ion programs at RHIC and LHC, suggesting the formation of a nearly perfect fluid whose evolution can be best described within the framework of relativistic viscous hydrodynamics in two and three dimensions, properly accounting for a causal and stable relativistic treatment of dissipative effects quantified through the ratio of shear viscosity to entropy density ($\eta/s$), to second order in velocity gradients~\cite{Baier:2006um,Baier:2006gy,Song:2007ux,Heinz:2009cv}. In such macroscopic model calculations, the $\eta/s$ ratio is introduced as a free input parameter in the numerical code describing the hydrodynamic evolution of the fireball. Its value is determined to best match the experimental data. However, coupling the viscous fluid dynamic description of the high-temperature plasma phase with a microscopic Boltzmann simulation of the late hadronic phase yields an improved description of the data~\cite{Song:2010aq,Ryu:2012at}. Several different formulations of these hybrid models are available in the literature. The VISHNU model, which incorporates a boost-invariant longitudinal velocity profile, matches the $(2 + 1)$-dimensional viscous hydrodynamic algorithm VISH2+1~\cite{Song:2007fn,Song:2007ux,Shen:2011kn} for solving the second-order Israel–Stewart equations~\cite{Israel:1979wp} to the microscopic UrQMD transport model~\cite{Bass:1998ca,Bleicher:1999xi} used as an afterburner. In contrast, the MUSIC+UrQMD model~\cite{Ryu:2012at} adopts a full $(3 + 1)$-dimensional expansion and uses the Israel–Stewart formalism for viscous evolution. By employing IP-Glasma initial conditions~\cite{Schenke:2012wb,Schenke:2012hg}, this model accurately describes the centrality and $p_{T}$ dependence of charged hadron flow harmonics, with an average $\eta/s = 0.2$ for Pb-Pb collisions at the LHC and a relatively smaller value of $\eta/s = 0.12$ for Au-Au collisions at top RHIC energy~\cite{Gale:2012rq}. Early fluid-dynamical simulations were mostly focused on shear viscosity as the dominant source of dissipation of the QGP fluid produced in heavy-ion collisions. However QCD being a non-conformal theory, for typical temperatures of the fireball produced in nuclear collisions at RHIC and LHC, the bulk viscosity should also be non-zero and may become large enough to influence the evolution of the medium. While shear viscosity primarily affects collective behaviour and momentum anisotropy, bulk viscosity retards the overall rate of radial expansion and hence reduces average momentum of the produced particles. In literature hydrodynamic simulations of heavy-ion collisions including both the shear and bulk viscous corrections have been performed with various initial conditions~\cite{Monnai:2009ad,Song:2009rh,Bozek:2009dw,Denicol:2009am,Denicol:2010tr, Roy:2011pk, Dusling:2011fd, Bozek:2012fw, Noronha-Hostler:2013gga, Rose:2014fba, Ryu:2015vwa, Ryu:2017qzn}. The studies demonstrate that bulk viscosity can have a visible effect on the bulk observables of heavy-ion collisions at RHIC and LHC. Extracted value of $\eta/s$ depends on the corresponding value of bulk viscous coefficient ($\zeta/s$). Simulations of Au-Au collisions at RHIC~\cite{Bozek:2009dw}, using Glauber model initial conditions have shown that a moderate value of bulk viscosity $\zeta/s = 0.03-0.04$ correspond to a relatively small value of shear viscosity $\eta/s = 0.1$ and can reasonably describe the mass ordering of $v_{2}$ of identified hadrons without spoiling the agreement in the HBT radii and $p_{T}$ spectra. Simulations with MUSIC hydrodynamical model with IP-Glasma initial conditions~\cite{Rose:2014fba}, have shown that inclusion of a finite bulk viscosity coefficient leads to a better description of the integrated ansiotropic flow coefficients ($v_{n}$) in in ultra-central Pb-Pb collisions at LHC, while reducing $\eta/s$ by $32 \%$ from $0.22$ to $0.15$. An improved agreement is also observed between the realistic hybrid simulations~\cite{Ryu:2015vwa, Ryu:2017qzn} using MUSIC+UrQMD model with IP-Glasma initial conditions and a multitude of integrated and differential measurements from Pb-Pb collisions at LHC and Au-Au collisions at RHIC, in presence of bulk viscosity. Inclusion of finite bulk viscosity of the order $\zeta/s \simeq 0.3$ near the QCD phase transition, can simultaneously describe the centrality dependence of multiplicity and $<p_{T}>$ of pions, kaons and protons and reduces the shear viscosity by almost $50 \%$ in order to preserve the model agreement with the measured harmonic flow coefficients. In short state of the art hydrodynamic description of heavy-ion collisions need to consider both shear and bulk viscosities during the fireball evolution for simultaneous description of experimentally observed radial flow and azimuthal anisotropy of the final state hadrons. Recently Bayesian analyses~\cite{Bernhard:2019bmu, Nijs:2020ors, JETSCAPE:2020shq, JETSCAPE:2020mzn} have also been performed for a quantitative estimation of the specific shear and bulk viscosities of the strongly interacting matter produced in relativistic heavy-ion collisions.

In the pre-LHC era, around a time  when the formulation of the relativistic viscous hydrodynamical models for heavy-ion collisions were still at a developing stage, several phenomenological models were proposed in literature to estimate  $\eta/s$ from experimental data on collective flow~\cite{Drescher:2007cd,Masui:2009pw,Tang:2009vp,Lacey:2006bc}, transverse momentum fluctuations~\cite{Gavin:2006xd} and heavy quark transport~\cite{PHENIX:2006iih}. The approaches were in large part motivated by the conjectured lower bound of specific shear viscosity, for any relativistic thermal field theory: $\eta/s \ge {1 \over 4\pi}$, in natural units~\cite{Kovtun:2004de}. Estimated value of $\eta/s$ was reported to be within the range of 1-4 times of the KSS bound.

The approach adopted by the authors of Ref.~\cite{Drescher:2007cd} was based on the prescription of incomplete equilibration consideration introduced in Ref.~\cite{Bhalerao:2005mm}. Within this framework, departure from the ideal hydrodynamic behaviour was parameterized in terms of Knudsen number $K = \lambda/\bar{R}$. $\lambda$ denotes the mean free path of the constituents in system and $\bar{R}$ is the characteristic measure of the system size. Determination of $K$ is directly related to the viscosity of the fluid. Hydrodynamic limit is reached for sufficiently small values of $K$ ($K \rightarrow 0$), whereas larger values indicate  presence of non-trivial mean free path which leads to damping of momentum anisotropy and deviation from the ideal hydrodynamic behavior. Very small amount of shear viscosities, of the order of KSS bound was found to result a significant reduction in the predicted $v_{2}$~\cite{Teaney:2003kp}. Direct extraction of $K$ is possible from the centrality dependence of the measured charged particle $v_2$ scaled by the initial state spatial eccentricity ($\varepsilon$) and thus providing a  link between the transport properties of the medium and experimentally measured data. 

In Ref.~\cite{Drescher:2007cd} the authors used this formulation of incomplete thermalization to fully describe the dependence of $v_2$ on centrality and system-size in $\sqrt{s_{NN}} = 200$ GeV Au-Au and Cu-Cu collisions measured by PHOBOS collaboration at RHIC. Two different initial condition models namely a Glauber Monte Carlo (GMC) approach and a Color-Glass Condensate (CGC) approach were employed for estimation of initial state geometrical parameters. Both the models were found to describe the PHOBOS data reasonably well, though the transport coefficients and thermodynamic quantities extracted from the data were found to be sensitive to the underlying initial conditions. Even for most central Au-Au collisions, the measured $v_2$ was found to be at least $25 \%$ below the ideal hydrodynamic limit. The absence of perfect equilibration allowed for estimation of the effective inter-particle cross section in the produced fluid and of its shear viscosity in terms of $\eta/s$. For CGC initial conditions, the scaled flow extrapolated to $K \rightarrow 0$ was found to be lower than for GMC initial conditions by a factor 0.73, implying a large interaction cross section, lower viscosity and a softer equation of state (smaller speed of sound, $c_s$) for the former. The reported estimation of specific viscosity are $\eta/s=0.19$ for GMC initial conditions and $\eta/s=0.11$ for CGC initial conditions. Similar parametrization was employed in Ref.~\cite{Tang:2009vp}, to perform a two parameter fit for the centrality dependence of the eccentricity scaled elliptic flow of charged and identified hadrons in 200 GeV Au-Au collisions at RHIC. Depending on the initial conditions, for charged particles the system was seen to be $30 - 50 \%$ away from the ideal hydrodynamics. Identified hadrons exhibited a mass hierarchy with more saturation for heavier particles. Extracting medium temperature from the slope of the $m_{T}$ spectra of pions measured by STAR collaboration, the $\eta/s$ ratio was estimated as a function of centrality for GMC and CGC initial conditions. For both the cases, $\eta/s$ was found to be lower than that for He at $T_{c}$ with CGC initial conditions generating lower specific viscosity than that for GMC. Subsequently in Ref.~\cite{Chaudhuri:2010in}, the estimation of $K$ was generalized, by taking into account the entropy generation during the evolution of a viscous fluid. The centrality dependent charged particle $v_2$ data for Au-Au and Cu-Cu collisions $\sqrt{s_{NN}} = 62$ and 200 GeV as made available by PHOBOS collaboration were analyzed within the modified Knudsen ansatz. Within errors, data did not show any energy or system size dependence. For most central Au-Au collisions, $K^{-1}$ was seen to be around three times lower compared to isentropic expansion, while $v_2$ is around $40 \%$ below the ideal hydrodynamic limit. The specific shear viscosity is estimated to be $\eta/s = 0.17 \pm 0.1 \pm 0.2$. Similarity between the estimated values of $\eta/s$ between the two approaches (Knudsen ansatz and modified Knudsen ansatz) was attributed to the presence of viscous effects in the experimentally measured rapidity density ($dN/dy$) used for calculation of transverse areal density of the medium. In Ref.~\cite{Liu:2009zzo}, the authors studied the sensitivity of eccentricity scaled elliptic flow on the partonic cross section in semi central (impact parameter $b=8$ fm) Au-Au collisions at $\sqrt{s_{NN}} = 200$ GeV using a multiphase transport (AMPT) model. Variation of the simulated $v_2/\varepsilon$ of the partons with cross section was shown to follow the Knudsen number scaling, with generated partonic flow being $19 - 27 \%$ lower than its hydrodynamic limit. Hadronization of the partons through coalescence and the following hadronic dynamics (collision, decay) were seen to reduce the the extracted hydrodynamic limit by $20 \%$ and also reduce the deviation from the hydrodynamic limit by $26 \%$.

After beginning of the heavy-ion run at LHC, plethora of data on anisotropic flow measurements in heavy-ion collisions started becoming available from ALICE, CMS and ATLAS collaborations. Till date  ALICE has reported measurement of charged particle ansiotropic flow coefficients in $\sqrt{s_{NN}} = 2.76$ and 5.02 TeV Pb-Pb and 5.44 TeV Xe-Xe collisions. CMS has also published charged particle $v_2$ data as a function of collision centrality in $\sqrt{s_{NN}} =2.76$ and 5.02 TeV Pb-Pb collisions. Results on centrality dependence of $v_{n} (n=2-6)$ have also been published by ATLAS collaboration, in the central region ($|\eta| < 2.5$), for six different $p_{T}$ ranges up to 20 GeV/$c$. In Ref.~\cite{Korotkikh:2014rma} the authors used then available CMS $v_2$ data in $\sqrt{s_{NN}} =2.76$ TeV Pb-Pb collisions to extract effective partonic cross section and the average QGP $\eta/s$ ratio within the incomplete equilibration scenario, with GMC initial conditions. The extracted value of effective inter-particle cross section was $\sigma = 3.1 \pm 0.2$ mb which reported to yield $\eta/s = 0.17 \pm 0.02$ of the produced fireball. However the calculations contain some serious conceptual and calculation flaws in evaluation of $\eta/s$. For estimation of $K^{-1}$, the corresponding transverse areal density is calculated from the measured charged particle pseudo-rapidity density ($dN/d\eta$) at midrapidity. However while using the extracted $\sigma$ for calculating $\eta/s$ the corresponding fluid density is evaluated from $\rho = \frac{g}{\pi^{2}}T^{3}$, a formula valid for a purely classical gluonic gas, with $g=16$ as the gluon degeneracy, whereas the temperature was taken as the chemical freeze out temperature $T=166$ MeV. More importantly using this prescribed values of density and temperature and extracted value of $\sigma$, $\eta/s$ comes out to be $~0.88$ instead of $~0.17$ as quoted by the authors. To the best of our knowledge, no other analysis of the available LHC data within the incomplete thermalization framework has been carried out yet. Also none of the studies employing eccentricity scaling of elliptic flow within the incomplete thermalization model have ever attempted to extract the specific bulk viscosity of the nuclear media created in heavy-ion collisions. 
 Though over the years the formulation of relativistic dissipative hydrodynamics containing the coefficients of both shear and bulk viscosity have reached to a mature level, still it would be interesting to test the incomplete thermalization scenario in the context of the available LHC data collected at much higher collision energies as compared to RHIC. In the present article we thus perform a reexamination of the latest charged particle $v_2$ data of Pb-Pb and Xe-Xe collisions, as available from ALICE and CMS collaborations at LHC, within the incomplete equilibration framework. The measured $v_{2}$ as a function of collision centrality, reported by ATLAS collaboration is integrated over $p_{T}$ up to 60 GeV/c\cite{ATLAS Collaboration} and thus has significant non-thermal contribution. Hence we have not used this result in our analysis. For completeness we have also reanalyzed the centrality dependence of integrated $v_2$ data from PHOBOS collaboration in $\sqrt{s_{NN}}=200$ GeV Au-Au and Cu-Cu collisions at RHIC. For all the above cases, we have consistently used $v_{2}$ values obtained from two particle correlation with $p_{T} < 5.0$ GeV/$c$.

The organization of rest of the article is the following. In section~\ref{methodology} we briefly describe the phenomenological construct of the incomplete thermalization framework. The required initial condition models as employed in the present work are also discussed in brief. Analysis results are depicted in section~\ref{results}. Finally we summarize and conclude in section~\ref{summary}. 

 \section{Incomplete Thermalization Model}
 \label{methodology}

\begin{figure*}
    \centering
    \includegraphics[width=0.7\textwidth]{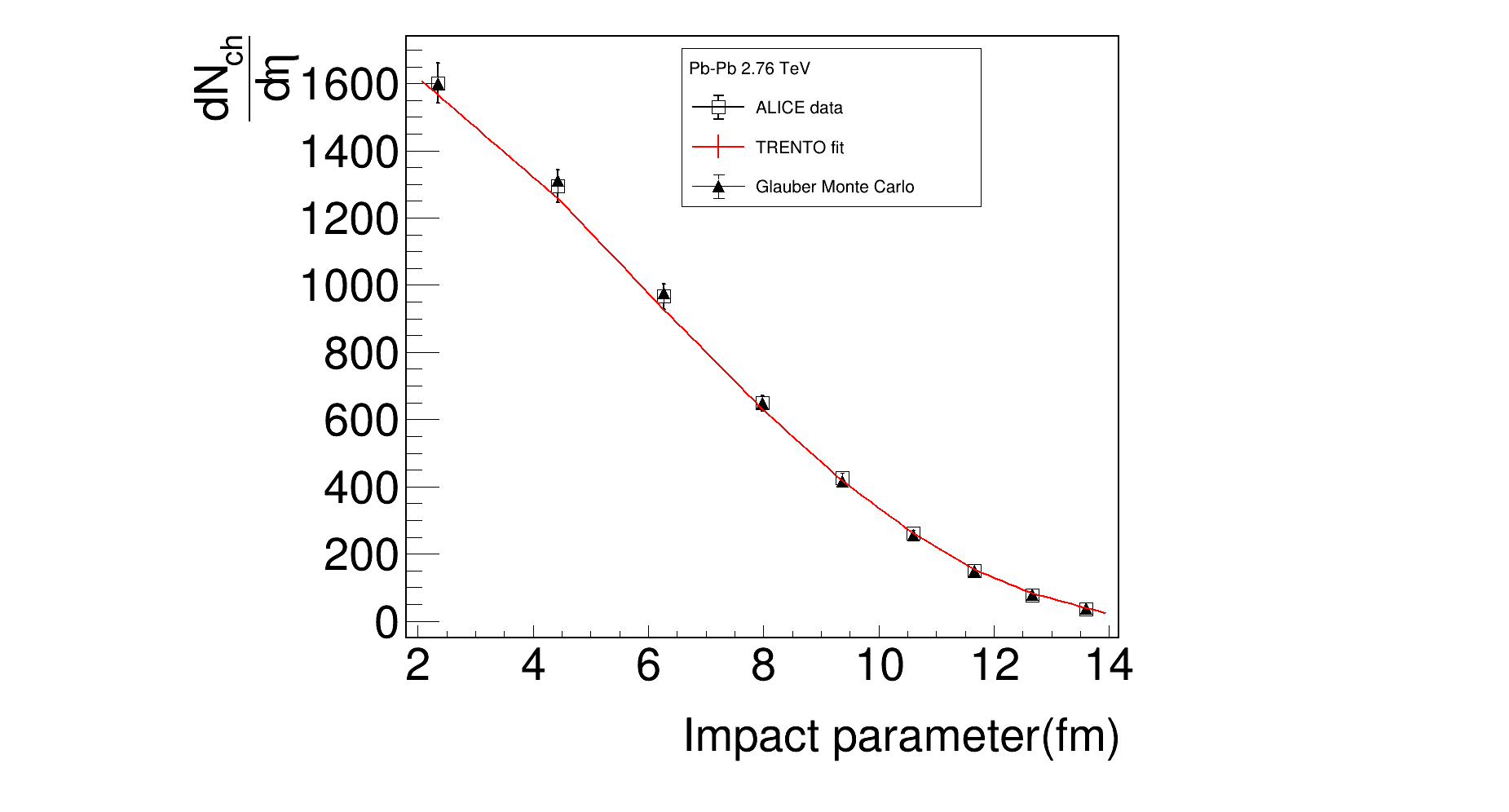}
    \caption{Comparison of charged particle pseudorapidity density obtained from different initial conditions models with ALICE data in $\sqrt{s_{NN}}=2.76$ TeV Pb-Pb collisions. The ALICE data points and Glauber Monte Carlo (GMC) points are shown on top of each other. The TRENTO results are shown as a fit to the data points in order to estimate the proportionality constant for conversion of integrated reduced thickness to charged particle pseudorapidity density.}
    \label{multi_check}
\end{figure*}

The scale invariance of ideal hydrodynamics demands, that in non-central collisions, the generated $v_2$ in final state is proportional to the initial state spatial eccentricity ($\varepsilon$) of the asymmetric overlap region~\cite{Bhalerao:2005mm}. If the re-scatterings among the constituents in the overlap zone are strong enough to drive the produced medium to perfect equilibrium then it behaves like an ideal fluid and the ratio $v_2 \over \varepsilon$ should be a constant independent of shape and size of the system, with a magnitude that depends on the underlying Equation of State (EoS). Any deviation from this perfect scaling of $v_2$ would indicate presence of dissipative effects in the medium and the produced fluid is viscous. Hence the ratio $v_2 \over \varepsilon$ as a function of particle density serves as a sensitive gauge to test if the medium produced in nuclear collisions approaches ideal hydrodynamic behaviour. In general the scaling ratio $\frac{v_n}{\varepsilon_n}$ is likely be affected by other eccentricities than that of order $n$ owing to the non-linear coupling. But for $n=2$ this effect is rather small \cite{Bhalerao:2011yg}. 
In the incomplete thermalization framework, the eccentricity scaled elliptic flow is parameterized  as
 \begin{equation}
   \frac{v_{2}}{\varepsilon} = \frac{v^{hydro}_{2}}{ \varepsilon}\: \frac{1}{1+ \frac{K}{K_{0}}}
  \label{v2_scale}
 \end{equation}

 where $v_{2}^{hydro}$ is interpreted as the hydrodynamic limit of the elliptic flow. ${(1+ \frac{K}{K_0})}^{-1}$ is the dissipative correction factor, where $K_0$ is assumed to be a constant to be determined by fitting this formula to data. $K$ on the other hand denotes the Knudsen number, a dimensionless parameter characterizing the degree of thermalization of the produced medium. It is defined as 
\begin{equation}
 K^{-1} =\frac{\bar{R}}{\lambda}
 \label{Knu}
\end{equation}
where $\bar{R}$ denotes the transverse size of the system and $\lambda ={1/{\rho\sigma}}$ is the mean free path of the constituents. Assuming isentropic evolution of the fireball, the particle number density $\rho$ at the initial stage, due to predominant longitudinal expansion decreases as the inverse of proper time $\tau$~\cite{Bjorken:1982qr}:

 \begin{equation}
   \rho(\tau) = \frac{1}{\tau S_T}\frac{dN}{dy}
  \label{den}
\end{equation}

 where $dN/dy$ denotes the total hadron (charged +neutral) multiplicity per unit rapidity and can be estimated from the charged particle pseudorapidity density $\frac{dN_{ch}}{d\eta}$ using $\frac{dN}{dy} = 1.5*1.15*\frac{dN_{ch}}{d\eta}$ \cite{PHOBOS Collaboration:0711.3724}. The multiplicative factors are to account for the neutral particles and to convert pseudorapidity density to rapidity density respectively. $S_{T}$ is the transverse overlap area between two colliding nuclei and can be defined as

 \begin{equation}
   S_{T} = 4\pi\sqrt{\sigma_{x}\sigma_{y}}
 \label{area}
   \end{equation}
   $\sigma_{x}$ and $\sigma_{y}$ are the standard deviations of the participant positions along x and y axis respectively.
 One may note that, $K$ should be evaluated at a characteristic timescale $\tau = \bar{R}/c_{s} $~\cite{Bhalerao:2005mm}, required for build up of $v_{2}$, when all parts of the system are informed about the initial spatial anisotropy such that,

\begin{equation}
  \frac{1}{K} = \frac{1}{S_{T}} \frac{dN}{dy} \sigma c_{s}
  \label{invK}
 \end{equation}

where $c_s$ is the speed of sound in the medium. Finally the eccentricity of the overlap zone ($\varepsilon_{2}$),  used to characterize the initial collision geometry can be defined as: 
\begin{equation}
  \varepsilon_{2} = - \frac{\langle r^2 e^{i2\phi}  \rangle}{\langle r^2 \rangle}
  \label{epsilon}
 \end{equation}

In order to apply the Knudsen number ansatz to the heavy-ion data we now need a suitable initial condition model to estimate the geometric quantities defined above. For our present study we use the results from two initial condition models namely the most widely used two component Glauber Monte-Carlo (GMC) model and the more recently proposed Reduced Thickness Event-by-event Nuclear Topology model (TRENTO). Both are non-dynamical models which neglect pre-equilibrium dynamics and generate static profiles of energy or entropy density distribution directly at the thermalization time of the medium. 

The Glauber model is a multiple-collision model in which a nucleus-nucleus collision is treated as an independent series of nucleon-nucleon collisions. To implement Glauber model for the present work, we utilize the foundation of a published Glauber MC software package TGLAUBER MC~\cite{tgmc}, originally developed by the PHOBOS Collaboration at RHIC~\cite{Alver:2008aq} and later updated~\cite{Loizides:2014vua} to accommodate the collision of deformed nuclei. The position of the nucleons inside the nucleus are determined stochastically according to probability density modeled from nuclear charge densities extracted in low energy electron scattering experiments. A minimum inter-nucleon separation between the centres of the nucleons are commonly required to determine the nucleon positions inside the given nucleus. The two nuclei are collided assuming the nucleons travel in a straight line along the beam axis (eikonal approximation) such that nucleons are tagged as wounded (participant) or spectator. In Eq.~\ref{epsilon} the averages are over the participant distribution.  The inelastic nucleon-nucleon cross section ($\sigma_{NN}$) is used to determine how close the nucleon trajectories need to be to interact and extracted from a fit to the existing data for total and elastic cross sections in p-p collisions. It is only a function of the collision energy and we use a value of $\sigma_{NN} =42$ mb for $\sqrt{s_{NN}}=200$ GeV, $\sigma_{NN} =64$ mb for $\sqrt{s_{NN}}=2.76$ TeV and  $\sigma_{NN} =72$ mb for $\sqrt{s_{NN}}=5.02$ TeV in our calculations. To parameterize the initial entropy deposition the model employs a two component ansatz in terms of the number of participant nucleons and the number of binary collisions. Following this, the charged particle pseudorapidity density can be estimated as 

\begin{equation}
  \frac{dN_{ch}}{d\eta}= n_{pp} [ (1-x)N_{part} \; + \; xN_{coll}]
  \label{dndy}
 \end{equation}
where $N_{part } $ denotes the number of participants and $N_{coll}$ is the number of binary collisions. Here $x$ represents the contribution of hard collisions to the particle production. $n_{pp}$ is the multiplicity corresponding to p-p collision at the same energy. In the present work, we have used the relationship $n_{pp}= 2.5-0.25(log(\sqrt{s_{NN}})^2)+ 0.023(log(\sqrt{s_{NN}})^2)^2$ to estimate $n_{pp}$.~\cite{Kharzeev:2000ph}. Two component Glauber ansatz was found to be inconsistent with the results from ultra-central U-U collisions at RHIC, which showed the violation of the scaling particle production with the number of binary collisions.

TRENTO~\cite{Moreland:2014oya} on the other hand, is a relatively new parametric initial-condition model for high energy nuclear collisions based on eikonal entropy deposition via a reduced thickness function. It is an effective model which generates realistic initial entropy profiles without assuming any specific physical mechanism for entropy production, pre-equilibrium dynamics or thermalization. The model is found to simultaneously describe the measured multiplicity distributions in p-p, p-A and A-A collisions and generate eccentricity harmonics in nuclear collisions consistent with experimental flow constraints. It uses a generalized mean ansatz to parameterize the entropy created at mid rapidity at the hydrodynamic thermalization time. The initial entropy density (consequently the charged particle pseudorapidity density) in this model is proportional to the reduced thickness function, which itself is a generalized mean of fluctuated participant thickness functions and is given by

\begin{equation}
{T_{R}(p;T_{A},T_{B})} \equiv {\left({{T^{p}_{A} +T^{p}_{B}} \over 2} \right)^{1/p}}
\label{TR}
\end{equation}

where $T_{A,B}$ denote the participant thickness of the two colliding nuclei. The dimensionless parameter $p$ takes different values for different physical mechanisms for entropy generation. For $p=1$, the reduced thickness is equivalent to a MC wounded nucleon model depositing a blob of entropy for each nucleon.  The eccentricity harmonics in this model are computed as weighted averages over the entropy profile (reduced thickness). A single roughly Gaussian profile of initial entropy deposition at the centre of the collision is adopted by the model for $p=1$. Negative values of $p$ suppresses the entropy deposition along impact parameter of the collision. Model-to-data comparison of charged particle multiplicity distribution at LHC showed that Pb-Pb collisions prefer a value $p=0$ over the others. For our present calculations we thus choose $p=0$. The shape parameter $k$, which accounts for multiplicity fluctuations, is set to unity. 

\begin{figure*}
    \centering
    \includegraphics[width=1.0\textwidth]{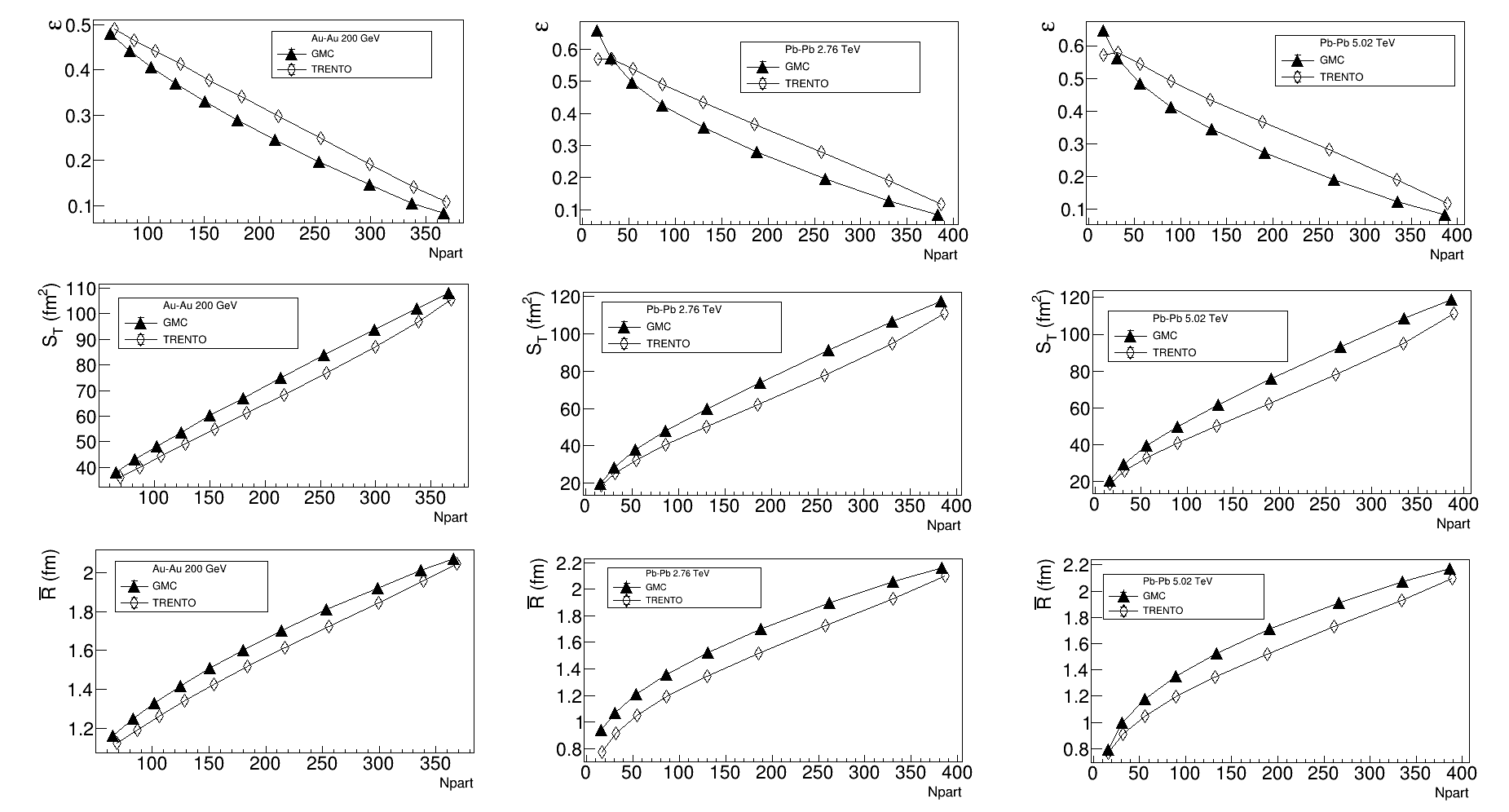}
    \caption{ The variation of eccentricity $\varepsilon \{2\}$ (top), transverse overlap area $S_{T}$ (middle) and transverse size $\bar{R}$ (bottom) as a function of number of participants using GMC and TRENTO initial conditions, for different collision systems.}
    \label{geomtric_quantities}
\end{figure*}

The essential difference between the two models lies in the fact that unlike Glauber approach, multiplicity in TRENTO is independent of the number of binary collisions due to the postulated scale invariant condition in the construction of reduced thickness ansatz. The charged particle pseudorapidity density $\frac{dN_{ch}}{d\eta}$ in TRENTO is obtained from the the integrated reduced thickness, which is proportional to the charged particle pseudorapidity density. Any experimental data at a given energy can be fit to set the constant of proportionality, and then it remains same for all systems and centrality classes at that energy. Both the two component ansatz of GMC and the integrated reduced thickness ansatz of TRENTO initial conditions are able to reproduce the centrality dependence of measured charged particle pseudorapidity density reasonably well, as shown in Fig.~\ref{multi_check}, where we have compared the model results with data from $\sqrt{s_{NN}}=2.76$ TeV Pb-Pb collisions measured by ALICE collaboration. Such model-to-data agreement is also seen at other collision energies. In case of GMC model, the fraction of binary collisions is set to $x = 0.12$. Before moving forward, it would be interesting to check the centrality dependence of the essential geometrical quantities namely $\varepsilon$, $S_{T}$ and $\bar{R}$ as obtained from the above two models of collisions. Results for the same are depicted in Fig.~\ref{geomtric_quantities} for $\sqrt{s_{NN}}=200$ GeV Au-Au and $\sqrt{s_{NN}}=2.76$ and 5.02 TeV Pb-Pb collisions. Centrality dependence is
expressed through $N_{part}$, the number of participating nucleons. While using the geometric mean of participant thickness functions to model initial entropy deposition, the resultant transverse profile of colliding nuclei is compressed, leading to smaller transverse size and transverse overlap area but larger eccentricity~\cite{Moreland:2019szz}. As a result, at all three investigated energies, eccentricity as obtained from TRENTO initial conditions are about 1.5 times larger than the eccentricity from GMC, for the more central collisions.

 \section{Analysis Results}
 \label{results}

\begin{figure*}{h}
    \centering
    \includegraphics[width=1.0\textwidth]{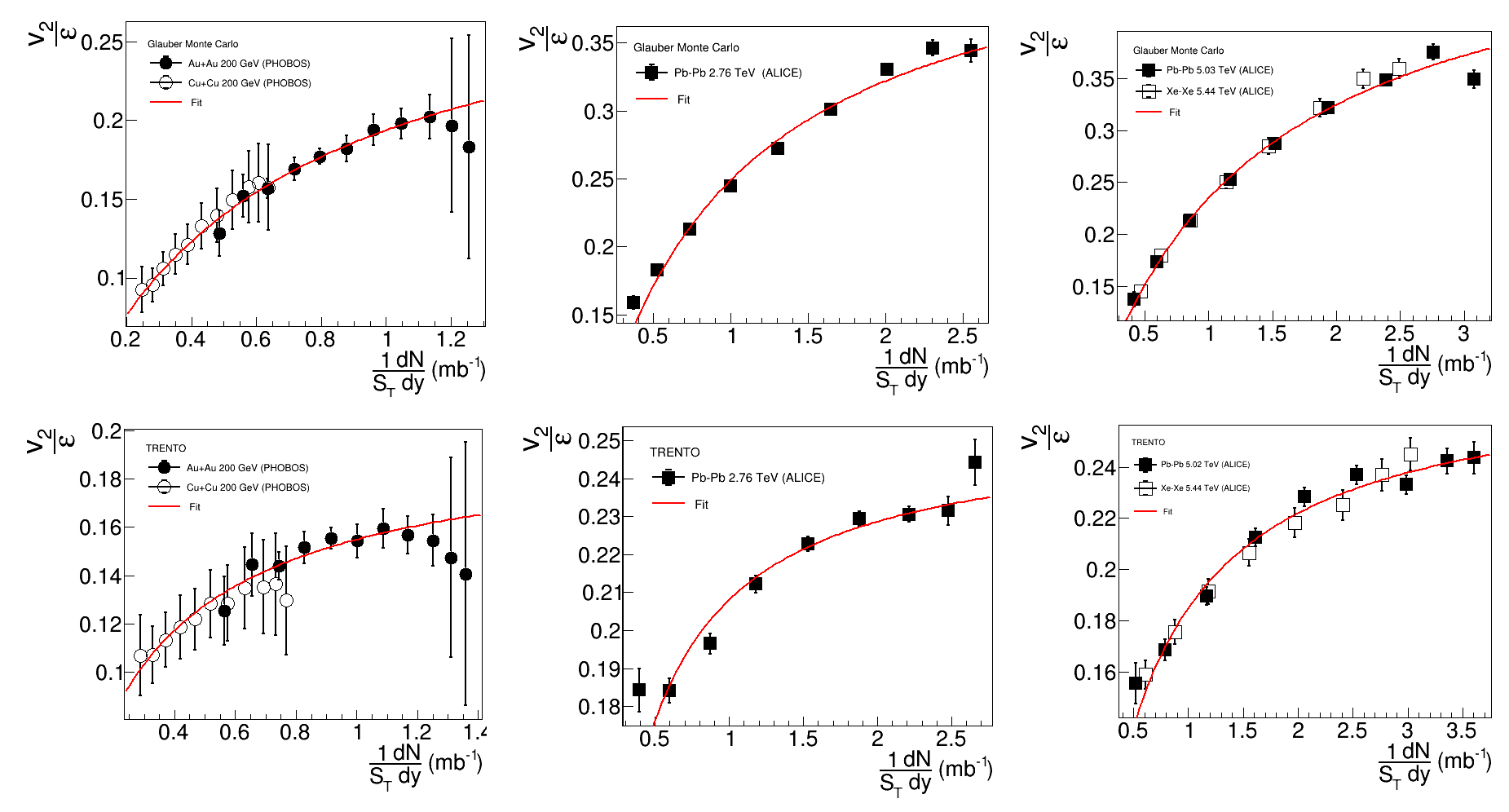}
    \caption{Variation of the eccentricity scaled elliptic flow with the transverse rapidity density for Au-Au and Cu-Cu collisions at $\sqrt{s_{NN}}=200$ GeV (PHOBOS), Pb-Pb collisions at  $\sqrt{s_{NN}}=2.76$ TeV (ALICE), Pb-Pb collisions at $\sqrt{s_{NN}}=5.02$ TeV and Xe-Xe collisions at $\sqrt{s_{NN}}=5.44$ TeV (ALICE) using GMC (top) and TRENTO (bottom) initial conditions. The GMC results are obtained for the two component fraction $x=0.12$ and  TRENTO results are obtained for the reduced thickness parameter $p=0$ and fluctuation parameter $k=1$.}
    \label{knudsenfit}
\end{figure*}

$\frac{dN}{dy}$, $S_{T}$ and $\varepsilon$ as obtained from either of the collision models, can now be used in Eq.~\ref{v2_scale} to fit the eccentricity scaled elliptic flow data ($v_{2} \over \varepsilon$) as a function of transverse rapidity density ($1/S_{T} dN/dy$) in Cu-Cu and Au-Au collisions at RHIC and in Pb-Pb and Xe-Xe collisions at LHC. The remaining unknown parameters are $K_{0}$ and $c_{s}$. Previous studies found that $K_{0}$ does not have a very significant effect on the extracted results as long as it is around unity. Following~\cite{Drescher:2007cd} we have used $K_{0} = 0.7$. On the other hand, $c_{s}$ has the maximum amount of uncertainty associated with it. It cannot be a constant for a rapidly evolving system as produced in relativistic heavy ion collisions, also it is a function of temperature of the system which itself rapidly changes during the expansion of the fireball.  Note that for the fit only the combination $K_{0}\sigma c_{s}$ is relevant. We have kept the fit parameter as $\sigma c_{s}$. The partonic cross section can be extracted by dividing this parameter by a suitable choice of $c_s$. Choosing a lower value of $c_{s}$ would result in a larger cross section ($\sigma $), as expected. But it has no effect on the extracted Knudsen number, which has been used to quantify the degree of equilibration within this model. The choice of $c_{s}$ also does not affect  the extracted  $\frac{v^{hydro}_{2}}{\epsilon}$. We now perform the fit employing the MINUIT package for $\chi^2$ minimization within ROOT analysis framework. The results of the model fit to the eccentricity scaled $v_2$ data using GMC and TRENTO initial conditions are shown in Fig.~\ref{knudsenfit} and corresponding fit parameters namely $\sigma c_s$ and $v^{hydro}_{2}/\epsilon$ are presented in Table~\ref{fitparams}. As mentioned before, at RHIC we have analyzed $v_{2}$ data from PHOBOS collaboration for $\sqrt{s_{NN}} = 200$ GeV Au-Au and Cu-Cu collisions~\cite{PHOBOS:2004vcu,PHOBOS:2006dbo}. The 90$\%$ C.L  systematic errors for Au-Au and Cu-Cu data have been converted to one standard deviation errors by dividing by 1.6.  At LHC $v_{2}$ data from ALICE collaboration for $\sqrt{s_{NN}} = 2.76$ TeV Pb-Pb collisions~\cite{ALICE:2010suc}, $\sqrt{s_{NN}} = 5.02$ TeV Pb-Pb collisions~\cite{ALICE:2016ccg} and $\sqrt{s_{NN}} = 5.44$ TeV Xe-Xe collisions~\cite{ALICE:2018lao} are analyzed. Charged particle $v_{2}$ values are integrated up to $5$ GeV/$c$. The systematic and statistical errors have been added in quadrature for all systems.


\begin{table}
  \caption{Effective inter-particle cross section times the speed of sound ($\sigma c_s$) and $v^{hydro}_{2} / \epsilon$ obtained from the model fit for GMC and TRENTO initial conditions, using data measured by PHOBOS collaboration at RHIC in $\sqrt{s_{NN}} = 200$ GeV Au-Au and Cu-Cu collisions, and by ALICE collaboration LHC, in $\sqrt{s_{NN}}=2.76$ TeV Pb-Pb, $\sqrt{s_{NN}}=5.02$ TeV Pb-Pb and $\sqrt{s_{NN}}=5.44$ TeV Xe-Xe collisions.}
  \centering
  \resizebox{245pt}{!}{
  \label{fitparams}
  \begin{tabular}{lllll}
    \hline\noalign{\smallskip}
      & \multicolumn{2}{l}{GMC} & \multicolumn{2}{l}{TRENTO}\\ 
      \hline\noalign{\smallskip}
      SYSTEM & $\sigma c_s$ (mb) & $\frac{v^{hydro}_{2}}{\epsilon}$  &  $\sigma c_s$ (mb) & $\frac{v^{hydro}_{2}}{\epsilon}$  \\
      \hline\noalign{\smallskip}
      Au-Au and Cu-Cu 200 GeV &  2.51 $\pm$ 0.45  &  0.31 $\pm$ 0.03 & 5.54 $\pm$ 1.22   &  0.19 $\pm$ 0.01  \\
      \hline\noalign{\smallskip}
      Pb-Pb 2.76 TeV & 1.70 $\pm$ 0.07 & 0.47 $\pm$ 0.01  & 5.19 $\pm$ 0.35 & 0.27 $\pm$ 0.01 \\
      \hline
      Pb-Pb 5.02 TeV and Xe-Xe 5.44 TeV & 1.26 $\pm$ 0.01 & 0.52 $\pm$ 0.01  & 2.62 $\pm$ 0.16 & 0.28 $\pm$ 0.01 \\
      \hline\noalign{\smallskip}
    \end{tabular}
  }
    
\end{table}

\begin{figure*}[!ht] 
 	\centering
 	\includegraphics[width=0.7\textwidth]{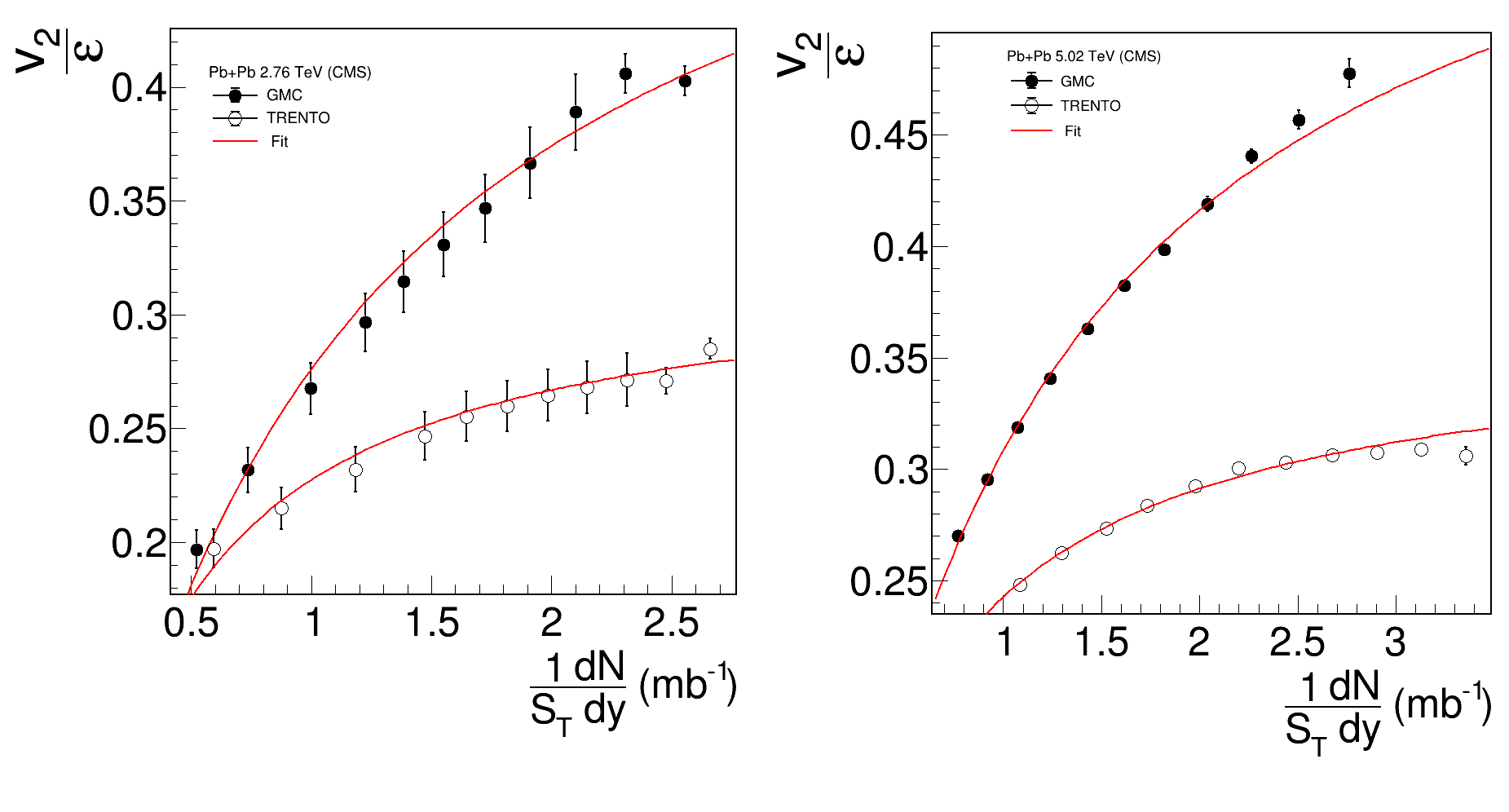} 
 	\caption{Variation of the eccentricity scaled elliptic flow with the transverse rapidity density for Pb-Pb collisions at $\sqrt{s_{NN}}=2.76$ TeV and 5.02 TeV measured by CMS collaboration, using GMC and TRENTO initial conditions. The GMC results are presented for the two component fraction $x=0.12$ and TRENTO results are presented for the reduced thickness parameter $p=0$ and fluctuation parameter $k=1$.}
 	\label{CMSfit}
 \end{figure*}

The values of the extracted $v_{2}^{hydro}/\varepsilon$ using GMC initial conditions indicate that the measured elliptic flow values are about 41\% lower than the ideal hydrodynamic limit for most central Au-Au collisions at $\sqrt{s_{NN}}=200$ GeV. Within uncertainty, estimated values of $\sigma$ and $v_{2}^{hydro}/\varepsilon$ are in agreement with the previous estimations at RHIC~\cite{Drescher:2007cd}. Pb-Pb collisions at $\sqrt{s_{NN}}=5.02$ TeV and Xe-Xe collisions at $\sqrt{s_{NN}}=5.44$ TeV are found to follow the same scaling. $v_{2}$ in most central Pb-Pb collisions  is about 31\% lower than the corresponding ideal hydrodynamic limit. The difference between the measured $v_2$ and the ideal hydrodynamic limit is found to be lowest for $\sqrt{s_{NN}}=2.76$ TeV Pb-Pb collisions and is about 26\%. For TRENTO initial conditions the differences become 28\% ,17\% and 10\% at at $\sqrt{s_{NN}}= 200$ GeV, 5.02 TeV and 2.76 TeV respectively. Similar analysis has also been performed with the available CMS data for Pb-Pb collisions at $\sqrt{s_{NN}}=$ 2.76 and 5.02 TeV~\cite{CMS:2012zex,CMS:2023bvg}, where $v_{2}$ values are integrated over $p_{T}$ up to 3 GeV/c. The fit results are displayed in Fig.~\ref{CMSfit} and corresponding fit parameters are reported in Table~\ref{CMS-params}. Note that the best fit parameter values obtained from ALICE and CMS data sets though similar but do not match with each other. This can be attributed to the difference in the measured value of $v_2$, which is because of the different  $p_{T}$  and $\eta$ ranges, used for the two experiments while extracting integrated $v_2$. The ALICE data corresponds to $|\eta| < 0.8$ and $p_T$ integrated up to 5 GeV/c whereas the CMS data corresponds to $|\eta| < 2.4$ and $p_T$ integrated up to 3 GeV/c. For all the analyzed data sets, it is unanimously observed that the extracted $\sigma c_s$ values from TRENTO initial conditions are a few times larger as compared to GMC initial conditions. The strong dependence of the extracted quantities on the choice of initial conditions is a consequence of the fact that the model takes many geometrical inputs ($S_T$, $\varepsilon$, $\bar{R}$), which are completely dependent on the underlying initial conditions for a given nucleus at a given energy. Larger $\varepsilon$ values from TRENTO cause the $v_{2}/\varepsilon$ to flatten faster, which translates to a larger $\sigma c_s$ from the fit. In Ref.~\cite{Drescher:2007cd} CGC initial conditions generating larger spatial eccentricity results in increased $\sigma$ and smaller hydrodynamic flow as compared to GMC initial conditions. On the other hand, in Ref.~\cite{Nagle:2009ip}, the authors examined in detail the present methodology of extraction of $K$ and $\eta/s$ using the PHOBOS data for centrality dependence of eccentricity scaled charged particle $v_{2}$ at RHIC. With GMC initial conditions, the spatial eccentricities were estimated for three different representative values of the binary collision fraction namely $x=0.0$, $0.13$ and $1.0$. Increased contribution of binary collisions resulted in larger initial eccentricities which in turn caused higher degree of flattening of $v_{2}/\varepsilon$ curve as a function of transverse density and thus larger $\sigma$ (for a constant value of $K_{0}c_{s}$) and smaller $v_{2}^{hydro}/\varepsilon$. As mentioned before in all our GMC calculations $x$ value is fixed to 0.12. However due to larger eccentricity values in TRENTO, we get larger $\sigma c_s$ and smaller $\frac{v^{hydro}_{2}}{\epsilon}$ at all three collision energies.
\begin{table}[ht]
	\caption{Effective inter particle cross section times the speed of sound ($\sigma c_s$) and $v^{hydro}_{2} / \epsilon$ obtained from the model fits using data measured by CMS collaboration at LHC.}
	\centering
	\resizebox{250pt}{!}{
		\begin{tabular}{lllll}
			\hline\noalign{\smallskip}
			& \multicolumn{2}{l}{GMC} & \multicolumn{2}{l}{TRENTO}\\ 
			\hline\noalign{\smallskip}
			SYSTEM & $\sigma c_s$ (mb) & $\frac{v^{hydro}_{2}}{\epsilon}$  &  $\sigma c_s$ (mb) & $\frac{v^{hydro}_{2}}{\epsilon}$  \\
			\hline\noalign{\smallskip}
			Pb-Pb 2.76 TeV & 1.41 $\pm$ 0.12 & 0.58 $\pm$ 0.02  & 3.23 $\pm$ 0.43 & 0.32 $\pm$ 0.01 \\
			\hline\noalign{\smallskip}
			Pb-Pb 5.02 TeV & 1.44 $\pm$ 0.03 & 0.64 $\pm$ 0.01  & 2.71 $\pm$ 0.81 & 0.36 $\pm$ 0.01 \\
			\hline\noalign{\smallskip}
		\end{tabular}
	}
	
	\label{CMS-params}
\end{table}

Our extracted $\sigma c_s$ values can now be plugged in Eq.~\ref{invK} to estimate the inverse Knudsen number in most central collisions at RHIC and LHC, to quantify the number of collisions a constituent of the produced medium undergoes. Corresponding results are tabulated in Table~\ref{K_eta}. The estimated values of $K^{-1}$, takes a minimum at $\sqrt{s_{NN}}=2.76$ TeV Pb-Pb collisions for both initial conditions. The Knudsen number so estimated can now also be used to estimate the specific shear viscosity ($\eta/s$). The authors in \cite{Drescher:2007cd,Nagle:2009ip} have calculated  $\eta/s$ , assuming the produced system to be a classical gas of massless particles with isotropic differential cross sections. For such a system $\eta=1.264 \frac{T}{\sigma}$. Using the relationship between entropy density ($s$) and particle density ($\rho$) for a classical ultra-relativistic gas $s=4\rho$, and using Eq.~\ref{den} we can write 
\begin{equation}
  \frac{\eta}{s} = 1.264 \frac{T}{4\rho\sigma}= 0.316 \frac{T}{\frac{\sigma c_s}{R S_T}\frac{dN}{dy}}
  \label{eta_over_s}
 \end{equation}

 \begin{figure*}[!ht]
    \centering
    \includegraphics[width=1.0\textwidth]{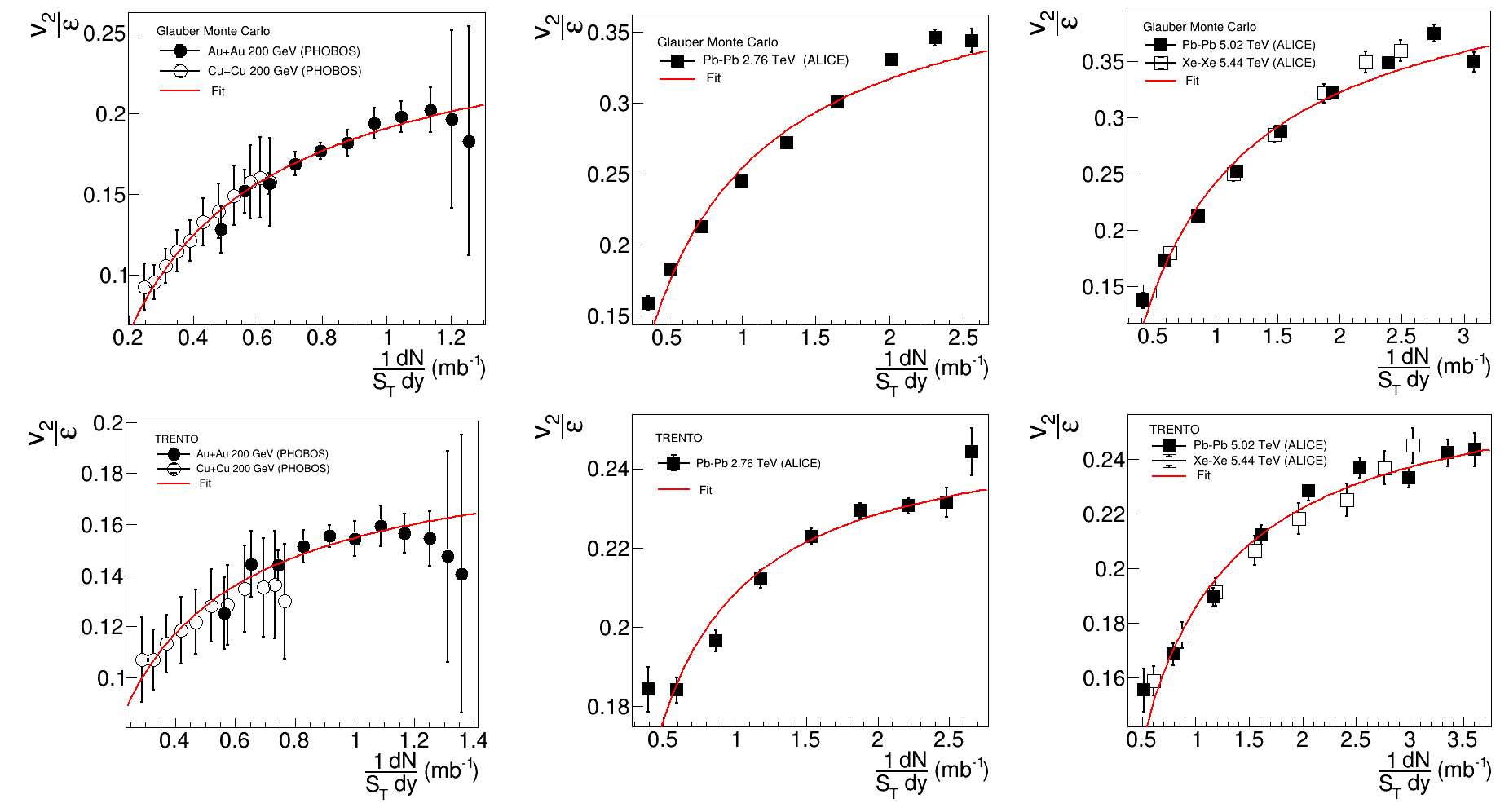}
    \caption{Variation of the eccentricity scaled elliptic flow with the transverse rapidity density using Pade approximant based model. The results are shown  for Au-Au and Cu-Cu collisions at $\sqrt{s_{NN}}=200$ GeV (PHOBOS), Pb-Pb collisions at $\sqrt{s_{NN}}=2.76$ TeV (ALICE) and  Pb-Pb collisions at  $\sqrt{s_{NN}}=5.02$ TeV and Xe-Xe collisions at $\sqrt{s_{NN}}=5.44$ TeV (ALICE) for GMC (top) and TRENTO (bottom) initial conditions. The GMC results are presented for the two component fraction $x=0.12$ and  TRENTO results are presented for the reduced thickness parameter $p=0$ and fluctuation parameter $k=1$. }
    \label{padefit}
\end{figure*}

 The value of $\eta/s$ is sensitive to the choice of the value of fireball temperature $T$. In previous calculations ~\cite{Drescher:2007cd, Nagle:2009ip} $T= 200$  MeV has been used for Au-Au collisions at $\sqrt{s_{NN}}$=200 GeV without any explicit explanation for the choice of the particular value. 
Assuming the medium to be a  classical ultra-relativistic gas also implies that the value of speed of sound, $c_s^2$ is $1/3$. 
Eq.~\ref{eta_over_s} can be augmented by using a more generalized form where the value speed of sound in the medium corresponds more closely to the estimated value of temperature of the system.  Instead of $s= 4\rho$ one can write the  more general expression $s=(1+ \frac{1}{c_s^{2}})\rho$. This reduces to $s= 4\rho$ when we assume non-interacting limit. Using this more general relation the specific viscosity can be expressed as
\begin{equation}
  \frac{\eta}{s} = 1.264 \frac{T}{(1+ \frac{1}{c_s^2})\rho\sigma}
  \label{eta_over_s_new}
\end{equation}

In previous calculations, as mentioned above, the authors have chosen ${c_{s}= \sqrt{1/3}}$, which corresponds to very high temperature limit. 
However, to compare different systems, we have chosen more realistic values for the speed of sound, using state of the art lattice QCD results~\cite{A. Bazavov} corresponding to effective temperature of the system using $T=\langle p_T\rangle/3.07$ \cite{Fernando G. Gardim}, where the mean transverse momentum $\langle p_T\rangle$ is the measured value for the corresponding systems\cite{STAR Colllaboration:044902,ALICE:R. Schicker,ALICE:5.02}. The values of  $\langle p_T\rangle$ and corresponding $T$ and $c_{s}^2$ values used in the present work for estimation of $\eta/s$ are tabulated in Table~\ref{T and Cs2}. 
\begin{table}{}
  \centering
   \caption{$K^{-1}$ and $\eta/s$ for the most central collisions(0-5 \%), obtained from the model fits for GMC and TRENTO initial conditions. The systematic uncertainty in $\eta/s$ corresponding arising from choice of $\bar{R}$(mentioned in subscript) and $T$(mentioned in superscript) are mentioned along with the errors from the fit.}
  \label{K_eta}
 \resizebox{250pt}{!}{
  \begin{tabular}{lllll}
 \hline\noalign{\smallskip}
  & \multicolumn{2}{l}{GMC} & \multicolumn{2}{l}{TRENTO}\\ 
    \hline\noalign{\smallskip}
    SYSTEM & $K^{-1}$ & $\eta/s$  &  $K^{-1}$ & $\eta/s$  \\
    \hline\noalign{\smallskip}
    Au-Au 200 GeV &  2.90  $\pm$ 0.52  & 0.15  $\pm$ $0.02^{+0.13}_{-0.02}$  &  7.52 $\pm$  1.66 & 0.06 $\pm$  $0.01^{+0.05}_{-0.01}$\\
    \hline\noalign{\smallskip}
    Pb-Pb 2.76 TeV & 4.00  $\pm$ 0.12 & 0.15 $\pm$ $0.01^{+0.11}_{-0.01}$ & 12.71 $\pm$ 0.85 & 0.05 $\pm$ $0.01^{+0.03}_{-0.01}$ \\
    \hline\noalign{\smallskip}
    Pb-Pb 5.02 TeV & 2.89$\pm$ 0.12 & 0.19 $\pm$ $0.01^{+0.14}_{-0.02}$  &  8.68 $\pm$ 0.52 & 0.07 $\pm$ $0.01^{+0.05}_{-0.01}$ \\
    \hline\noalign{\smallskip}
  \end{tabular}
  }
 
\end{table}
\begin{table}
  \caption{$\langle p_T \rangle$, corresponding temperature estimated using $T=\langle p_T\rangle/3.07$ and the corresponding $c_s^2$ from lattice QCD calculations for different collision systems}
  \centering
  \resizebox{245pt}{!}{
  \label{T and Cs2}
  \begin{tabular}{llll}
   
      \hline\noalign{\smallskip}
      SYSTEM &  $\langle p_T \rangle$ (GeV )& $T$ (GeV)  &  $c_s^2$    \\
      \hline\noalign{\smallskip}
      Au-Au and Cu-Cu 200 GeV &  0.577   &  0.188  & 0.202     \\
      \hline\noalign{\smallskip}
      Pb-Pb 2.76 TeV &  0.65 & 0.211  & 0.232  \\
      \hline
      Pb-Pb 5.02 TeV  & 0.681  & 0.220 & 0.241 \\
      \hline\noalign{\smallskip}
       \end{tabular}
   }
    
\end{table}

On the other hand, the transverse size of the system is calculated as $ \bar{R}= \sqrt{2(\langle r^2 \rangle - \langle r\rangle^2)}$ \cite{Fernando G}, where $r$ is the transverse coordinate of a participant.
The extracted values of $K^{-1}$ and $\eta/s$ as extracted for different systems have been shown in Table \ref{K_eta}. The symmetric errors around the central values of $K^{-1}$ and $\eta/s$ correspond to the fitting error in $\sigma c_{s}$ propagated following Eq.~\ref{invK} and Eq.~\ref{eta_over_s_new} respectively. With GMC initial conditions the central values of $K^{-1} (\eta/s)$ are smaller (larger) than the TRENTO initial conditions. This can be attributed to the larger $\varepsilon$ and hence larger $\sigma c_{s}$ values generated by TRENTO model as compared to the two component Glauber initialization. 


At this juncture it would be instructive to check the systematic uncertainties associated with Knudsen number formalism, due to variation of model parameters, for the extraction of $\eta/s$ from the experimental data. As mentioned earlier, in Ref.~\cite{Nagle:2009ip} such an exercise has been carried out in detail, for the $v_2$ measurements at RHIC. In the similar line we now study the sensitivity of our estimated results at LHC. To start with we note that in extraction of $\sigma$ (for fixed $K_{0}$ and $c_{s}$) any multiplicative scaling of the vertical axis quantity does not affect the fit parameters. However any multiplicative scaling of horizontal axis quantity $1/S_{T} dN/dy$ leads to a linear change in $\sigma$. For example, using a different multiplicative factor for transverse area $(S_{T} = 2\pi\sqrt{\sigma_{x}\sigma_{y}}$ or $S_{T}= \pi\sqrt{\sigma_{x}\sigma_{y}})$ has a non negligible effect on the extracted fit parameters. A larger multiplicative factor scales the cross section correspondingly though the results for Knudsen number and $\eta/s$ remain unchanged. The same is true if the multiplicative factor for the jacobian to convert from pseudorapidty to rapidity density is changed from 1.15 to 1.25.  However inclusion of fluctuation by including the cross term $\sigma_{xy}$ in Eq.~\ref{area} is seen to have a negligible effect on the resulting values of $S_{T}$. In literature, alternative formulation is also available for estimation of the transverse size of the system, $\bar{R}$, like the one prescribed in~\cite{Bhalerao:2005mm} where $\bar{R} = {1}/{\sqrt{\frac{1}{\langle{x^2}\rangle} + \frac{1}{\langle{y^2}\rangle}}}$
describes the size of the system along the gradient of matter distribution. We have found that using this prescription the values of transverse size and corresponding $\eta/s$ decreases by about $10\%$. Similarly we can also opt for a different estimation for the average temperature of the fireball.
For the present study, we have obtained average temperature  from a hydrodynamic model simulation\cite{Dasgupta:2018pjm}. The average is calculated over a time scale of $\tau = \bar{R}/c_{s}$. The estimated values of temperatures come out to be 233 MeV, 277 MeV and 291 MeV  for Au-Au collisions at 200 GeV, Pb-Pb collisions at 2.76 TeV and Pb-Pb collisions at 5.02 TeV respectively. These values are rather higher than the $T$ values extracted from the measured charged particle $p_{T}$ distributions (quoted in Table~\ref{T and Cs2}) and thus result in an increase in the corresponding value of $\eta/s$. Note such higher values of temperature would also imply an increase in corresponding $c_{s}^{2}(T)$. However any change in the absolute value of $c_{s}^{2}$ can be adjusted by the corresponding change in $\sigma$ to keep the fitted value of $\sigma c_{s}$ and hence extracted $\eta/s$ unchanged. The asymmetric systematic errors in $\eta/s$, originating from the variation of $\bar{R}$ and $T$ are also quoted in Table~\ref{K_eta}. As evident, for either of the initial state models, once the systematic uncertainties are taken into account, the extracted $\eta/s$ values show compatibility with the conjectured lower bound ($\sim0.08$) and more interestingly with the values extracted from full scale relativistic viscous hydrodynamic model simulations, which are more around $\eta/s \simeq 0.12$ The results also suggest the average $\eta/s$ of the fireball remains constant as a function of collision energy or temperature. 

To proceed further, it may be noted that Eq.~\ref{v2_scale} is not derived from any first principle theoretical argument, rather has been constructed to obey limiting functional forms corresponding to very large and very small values of $K$. The authors in~\cite{Nagle:2009ip} have thus suggested an alternative formulation of the functional relation between $v_2$ and $K$ based on Pade approximants, where the eccentricity scaled $v_{2}$ can be modeled as
 \begin{equation}
   \frac{v_{2}}{\epsilon} = \frac{v^{hydro}_{2}}{ \epsilon}\: \frac{1 + A(\frac{K}{K_{0}}) + B(\frac{K}{K_{0}})^{2} }{1+ C(\frac{K}{K_{0}})  + D(\frac{K}{K_{0}})^{2} + E(\frac{K}{K_{0}})^{3}}
   \label{pade}
  \end{equation}
 
 This equation abides to  the small and large $K$ limits as did the previous parametrization, if B = E and C = A + 1. It was seen that any arbitrary large value for $\sigma$ resulted an equally good description of the analyzed PHOBOS data with nearly same value of $\frac{v^{hydro}_{2}}{\epsilon}$ and an increasing value of the parameter $D$ for larger $\sigma$ values. The corresponding allowed range of $\eta/s$ was found to vary approximately from 0.03 to 0.2. The results indicate that any minimal modification of the Knudsen parametrization yields an order of magnitude amplification in the uncertainties in $\eta/s$. Our best fit results for RHIC and LHC data sets, with two opted initial conditions, are displayed in Fig.~\ref{padefit}. For both PHOBOS and ALICE data sets, the best fit corresponds to $D=2$, while we have kept the values of $K_{0}$ and $c_{s}$ same as before. We observe that this alternate parametrization yields cross section values that are more than two times larger corresponding to both the initial conditions as shown in Table~\ref{padefit-params}. This also translates to the extracted values of $K^{-1}$ and $\eta/s$ in Table~\ref{K-etaPade}. The authors in~\cite{Nagle:2009ip} observed an increase in the effective inter-particle cross section of about 25$\%$ only and a similar reduction in the hydrodynamic flow. In  our analysis, the deviation of measured $v_2$ from ideal hydrodynamics is found to be 33\% for Au-Au 200 GeV, 17\% for Pb-Pb 2.76 TeV and 21\% for Pb-Pb 5.02 TeV collisions for GMC initial condition whereas it is 27\% ,3\% and 11\% respectively for TRENTO initial conditions. We observe that the decrease in $\frac{v^{hydro}_{2}}{ \epsilon}$ is about 13$\%$, 9$\%$ and 12$\%$ for Au-Au 200 GeV, Pb-Pb 2.76 TeV and Pb-Pb 5.02 TeV  collision systems respectively. 

 \begin{table}[ht]
   \caption{Effective inter-particle cross section ($\sigma$) and $v^{hydro}_{2} / \epsilon$ obtained from the Pade approximant based model fit for GMC and TRENTO initial conditions}
  \centering
  \resizebox{250pt}{!}{
    \begin{tabular}{lllll}
      \hline\noalign{\smallskip}
      & \multicolumn{2}{l}{GMC} & \multicolumn{2}{l}{TRENTO}\\ 
      \hline\noalign{\smallskip}
      SYSTEM & $\sigma c_s$ (mb) & $\frac{v^{hydro}_{2}}{\epsilon}$  &  $\sigma c_s$ (mb) & $\frac{v^{hydro}_{2}}{\epsilon}$  \\
      \hline\noalign{\smallskip}
      Au-Au  and Cu-Cu 200 GeV & 7.48  $\pm$ 0.76   & 0.27 $\pm$  0.01 &  12.24$\pm$ 2.28   &  0.19 $\pm$ 0.01\\
      \hline\noalign{\smallskip}
      Pb-Pb 2.76 TeV & 4.83 $\pm$0.12  & 0.42 $\pm$ 0.01  & 11.04 $\pm$ 0.69 & 0.27 $\pm$ 0.01 \\
      \hline\noalign{\smallskip}
      Pb-Pb 5.02 TeV  and Xe-Xe 5.44 TeV & 3.80 $\pm$ 0.1  & 0.46 $\pm$ 0.01  & 5.80 $\pm$ 0.30 & 0.29 $\pm$ 0.01 \\
      \hline\noalign{\smallskip}
    \end{tabular}
  }
  \label{padefit-params}
\end{table}
 
 \begin{table}[ht]
   \caption{ $K^{-1}$ and $\eta/s$ for the most central Au-Au and Pb-Pb collisions at RHIC and LHC, obtained from the pade approximant based model fit for GMC and TRENTO initial conditions.The systematic uncertainty in $\eta/s$ corresponding arising from choice of $\bar{R}$(mentioned in subscript) and $T$(mentioned in superscript) are mentioned along with the errors from the fit. }
  \centering
   \resizebox{250pt}{!}{
   \begin{tabular}{lllll}
      \hline\noalign{\smallskip}
      & \multicolumn{2}{l}{GMC} & \multicolumn{2}{l}{TRENTO}\\ 
      \hline\noalign{\smallskip}
 
    SYSTEM & $K^{-1}$ & $\eta/s$  &  $K^{-1}$ & $\eta/s$  \\
    \hline\noalign{\smallskip}
    Au-Au 200 GeV &  8.64 $\pm$ 0.88 & 0.05 $\pm$ $0.01^{+0.04}_{-0.01}$  &  16.62 $\pm$  3.10 & 0.03 $\pm$  $0.01^{+0.02}_{-0.003}$\\
    \hline\noalign{\smallskip}
    Pb-Pb 2.76 TeV & 11.35  $\pm$ 0.29 & 0.05 $\pm$ $0.01^{+0.04}_{-0.01}$   & 27.00 $\pm$ 1.68 & 0.02 $\pm$ $0.01^{+0.02}_{0.002}$ \\
    \hline\noalign{\smallskip}
    Pb-Pb 5.02 TeV & 8.7 $\pm$ 0.2 & 0.06 $\pm$ $0.01^{+0.05}_{-0.01}$  &  19.24 $\pm$ 0.99 & 0.03 $\pm$ $0.01^{+0.02}_{-0.003}$ \\
    \hline\noalign{\smallskip}
  \end{tabular}
  }
   \label{K-etaPade}
\end{table}

The central values for the extracted $\eta/s$ can be seen to go well below the conjectured lower bound of $1/4\pi$ for TRENTO initial conditions. This appears to be a cumulative effect of the shrunken energy density profile of colliding nuclei in the TRENTO model, and the higher values of the parameter $\sigma c_s $ from the fit using this alternate ansatz. The values of $\eta/s$  for GMC initial conditions are still around the conjectured lower bound after accounting for the systematic uncertainty. As the authors in~\cite{Nagle:2009ip} have pointed out, the current formalism does not constrain the $\eta/s$ to any quantum bound. As such there is no first principle argument to constrain the $\eta/s$ for the produced medium  to have a lower bound of $1/4\pi$, but if such a bound exists, then the ideal hydrodynamic limit cannot be reached.
 




  

 


In Ref.~\cite{Chaudhuri:2010in} another alternative ansatz was suggested to incorporate the entropy generation during the evolution of the fireball, in estimation of Knudsen number, arising from the shear viscosity. In this ansatz the Knudsen number is parameterized as 

 \begin{equation}
     \frac{1}{K}= \sigma c_{s}\frac{1 dN}{S_{T} dy} \left[ 1 + \frac{2}{3\tau_{i}T_{i}} \left( \frac{\eta}{s}\right)   \right]^{-3}
  \end{equation}
which modifies the expression for eccentricity scaled elliptic flow to 

\begin{equation}
     \frac{v_{2}}{\varepsilon}= \frac{v^{hydro}_{2}}{\varepsilon} \frac{ \left[ 1 + \frac{2}{3\tau_{i}T_{i}} \left( \frac{\eta}{s}\right)   \right]^{-3}}{\frac{1}{K_{0}\sigma c_{s}} + \left[ 1 + \frac{2}{3\tau_{i}T_{i}} \left( \frac{\eta}{s}\right)   \right]^{-3} }
  \end{equation}
where $T_{i}$ denotes the temperature of the fireball at the time scale $\tau_{i}$. Note that the above analytical form of viscous correction is based on dissipative effects to first order in gradient expansion and suffers from nonphysical reheating artifacts at early times~\cite{Baier:2006um}. For completeness, we have applied this modified Knudsen ansatz to the $v_{2}$ data in Pb-Pb collisions at $\sqrt{s_{NN}} =2.76$ TeV with GMC initial conditions. $\sigma$ needs to be fixed by hand in this formalism. We have checked that to get the similar values for $K^{-1}$ and $\eta/s$ as reported in Ref.~\cite{Chaudhuri:2010in} we need to fix the $\sigma$ to 6.5 mb for $c_s^2 =1/3$.  We have also observed that in this formalism, increasing the cross section by hand yields a larger values of the parameter $\frac{1}{\tau_{i}T_{i}}(\frac{\eta}{s})$, possibly indicating early thermalization. It may be interesting to note that the Knudsen parametrization has also been studied within the framework of relativistic viscous hydrodynamics~\cite{Masui:2009pw} with two different EoS and two different values of $\eta/s$. The effect of EoS is clearly visible in the magnitude of $v_{2}/\varepsilon$, with harder EoS generating much larger value of ideal hydrodynamic limit. Value of $\eta/s$ is reflected by the change in the magnitude of $v_{2}/\varepsilon$ as a function of transverse density, with softer EoS corresponding to a smaller $\eta/s$.

So far we have discussed the extraction of $\eta/s$ from the eccentricity scaled $v_{2}$ data within the incomplete equilibration model framework and its heavy dependence on the model parameters. However as mentioned before modern hydrodynamics simulations have suggested that bulk viscosity also plays a crucial role in explaining the data from relativistic heavy-ion collisions. 
The model can be extended to estimate $\zeta/s$ by using results from kinetic theory. The presence of bulk viscosity leads to additional entropy production, as it tends to slow down the radial expansion of the produced medium and consequently affect the $p_{T}$ spectra and flow variables. The measured $v_{2}$ of course contains all the contributions, whether coming from shear and/or bulk viscosity which translate to deviation from perfect equilibration. The contribution of the specific bulk viscosity to the imperfect scaling of $v_{2/}\varepsilon$ can be estimated using the results from Ref.~\cite{Denicol:2014vaa}, which first computed the transport coefficients that couple the bulk viscous pressure to the shear stress tensor using the 14-moment approximation for a single component relativistic dilute gas as:

  \begin{equation}
     \frac{\zeta}{s}= 75 \frac{\eta}{s} \left( \frac{1}{3}-c_{s}^{2}\right)^{2}
  \end{equation}
  
  Using the $\eta/s$ value obtained earlier from GMC initial conditions and $ c_s^{2}=0.241$ (see Table~\ref{T and Cs2}) we arrive at a value of $\frac{\zeta}{s} \simeq 0.12$ for the most central Pb+Pb collisions at $\sqrt{s_{NN}} =$ 5.02 TeV. The corresponding value for TRENTO initial conditions is $\frac{\zeta}{s} \simeq 0.04$. Interestingly even if the above parametrization of $\zeta/s$ is linear in $\eta/s$, the bulk viscosity estimated within incomplete thermalization model does not necessarily show one to one correspondence with shear viscosity, due to its direct dependence on $c_{s}^{2}$. For example, instead of 220 MeV if we assume the fireball temperature to be 291 MeV, the $\eta/s$ for GMC model increases from 0.19 to 0.34 whereas $\frac{\zeta}{s}$ decreases to 0.08 due to an increase in $c_{s}^{2}$ from $0.241$ to $0.278$. As $c_{s}^{2} \rightarrow {1 \over 3}$ bulk viscosity will reduce to zero. 
At high temperatures, well above the QCD transition temperature, only $\eta/s$ has significant impact on $v_{2}$. As and when the system undergoes cooling and reach close to the transition temperature, the bulk viscosity starts to play a crucial role in determining the dynamics of the system.
  






\section{Summary and conclusion}
\label{summary}

In this article we make a quantitative reevaluation of the formulation for extraction of the Knudsen number ($K$), the specific shear viscosity ($\eta/s$) and the specific bulk viscosity ($\zeta/s$), within the incomplete thermalization scenario in relativistic heavy-ion collisions, realized via eccentricity scaling of the elliptic flow of the final state hadrons. Data on centrality dependent $v_{2}$ of unidentified charged hadrons measured in Pb-Pb and Xe-Xe collisions by ALICE and CMS collaborations at LHC are analyzed for this purpose, using GMC and TRENTO initial state models. The measured $v_{2}$ even for most central collisions Pb-Pb collisions at $\sqrt{s_{NN}} = 5.02$ TeV is below (ideal) "hydrodynamic limit". Sensitivity of the final results on the employed initial conditions are studied in detail. Impact of the Pade approximants based alternative formulation of the functional relation between $v_2$ and $K$,effect of entropy production during medium evolution are also investigated. The results highlight the importance of clarifying the initial state configurations as well as the ambiguities inherent in using a transport approach in a strongly coupled medium in the LHC energy regime. In general a larger spatial eccentricity from TRENTO is found to result in smaller deviation from the limiting ideal hydrodynamic flow and smaller value of $\eta/s$. We have also used the $\eta/s$ to calculate the $\zeta/s$ following a linear coupling between the transport coefficients. As none of the employed models includes evlution dynamics, the estiamted viscous coefficients represent effective values averaged over the temperature of the produced medium. Once the systematic errors associated with the extraction of $\eta/s$ is taken into account, the results appear consistent with a constant specific viscosity as a function of collision energy or fireball temperature, irrespective of the choice of the initial conditions. Moreover the estimated central values of $\eta/s$, particularly for TRENTO model lie below the conjectured lower bound for the collision systems under investigation. However within the associated systematic uncertainties the extracted specific shear viscosity appear to be consistent with relativistic viscous hydrodynamic model analyses which report a value around 0.12.

In conclusion, we thus note that the present study provides an intuitive picture of incomplete thermalization in terms of Knudsen number and presents a reasonable qualitative estimate of the viscous coefficients of the hot and dense medium produced in the heavy ion collisions at LHC. Even though the dependence of the extracted transport coefficients on the choice of initial conditions is not new but a quantitaive comparison of two different initial state models, within the incomplete equilibration framework using LHC heavy ion data has not been performed before. As compared to previous studies, we have updated the different model parameters ($T$, $c_{s}^{2}$, $\bar{R}$) as per our latest understanding of the heavy-ion data. Moreover our study corrects the flaws of previous analysis performed at LHC energy in estimating the shear viscosity. The incomplete thermalization model has also been applied for the first time to estimate the coefficient of bulk viscosity. With the inclusion of systematic errors our results are indeed consistent with the current understanding of viscous effects present in high energy heavy-ion collisions. Complex Bayesian analyses that use actual hydrodynamic simulations are rigorous and give  precise quantitative estimate of the viscous coefficients. However such simulations are numerically intensvive and demands heavy computation. In contrast the present work is based on a simple phenomenological model that gives rather qualitative but reasonable estimates of the transport coefficients of  hot and dense  nuclear matter that contemporary heavy ion community may find interesting.

\begin{acknowledgements}
We are thankful to Sushant K. Singh, Rupa Chatterjee, Rajeev S. Bhalerao and Amaresh Jaiswal for many stimulating discussions. 
\end{acknowledgements}


\begin{thebibliography}{99}

\bibitem{BRAHMS:2004adc}
I.~Arsene \textit{et al.} [BRAHMS],
Nucl. Phys. A \textbf{757}, 1-27 (2005)
doi:10.1016/j.nuclphysa.2005.02.130
[arXiv:nucl-ex/0410020 [nucl-ex]].

\bibitem{PHOBOS:2004zne}
B.~B.~Back \textit{et al.} [PHOBOS],
Nucl. Phys. A \textbf{757}, 28-101 (2005)
doi:10.1016/j.nuclphysa.2005.03.084
[arXiv:nucl-ex/0410022 [nucl-ex]].

\bibitem{STAR:2005gfr}
J.~Adams \textit{et al.} [STAR],
Nucl. Phys. A \textbf{757}, 102-183 (2005)
doi:10.1016/j.nuclphysa.2005.03.085
[arXiv:nucl-ex/0501009 [nucl-ex]].

\bibitem{PHENIX:2004vcz}
K.~Adcox \textit{et al.} [PHENIX],
Nucl. Phys. A \textbf{757}, 184-283 (2005)
doi:10.1016/j.nuclphysa.2005.03.086
[arXiv:nucl-ex/0410003 [nucl-ex]].

\bibitem{Voloshin:1994mz}
S.~Voloshin and Y.~Zhang,
Z. Phys. C \textbf{70}, 665-672 (1996)
doi:10.1007/s002880050141
[arXiv:hep-ph/9407282 [hep-ph]].

\bibitem{Baier:2006um}
R.~Baier, P.~Romatschke and U.~A.~Wiedemann,
Phys. Rev. C \textbf{73}, 064903 (2006)
doi:10.1103/PhysRevC.73.064903
[arXiv:hep-ph/0602249 [hep-ph]].

\bibitem{Baier:2006gy}
R.~Baier and P.~Romatschke,
Eur. Phys. J. C \textbf{51}, 677-687 (2007)
doi:10.1140/epjc/s10052-007-0308-5
[arXiv:nucl-th/0610108 [nucl-th]].

\bibitem{Song:2007ux}
H.~Song and U.~W.~Heinz,
Phys. Rev. C \textbf{77}, 064901 (2008)
doi:10.1103/PhysRevC.77.064901
[arXiv:0712.3715 [nucl-th]].

\bibitem{Heinz:2009cv}
U.~W.~Heinz, J.~S.~Moreland and H.~Song,
Phys. Rev. C \textbf{80}, 061901 (2009)
doi:10.1103/PhysRevC.80.061901
[arXiv:0908.2617 [nucl-th]].

\bibitem{Song:2010aq}
H.~Song, S.~A.~Bass and U.~Heinz,
Phys. Rev. C \textbf{83}, 024912 (2011)
doi:10.1103/PhysRevC.83.024912
[arXiv:1012.0555 [nucl-th]].

\bibitem{Ryu:2012at}
S.~Ryu, S.~Jeon, C.~Gale, B.~Schenke and C.~Young,
Nucl. Phys. A \textbf{904-905}, 389c-392c (2013)
doi:10.1016/j.nuclphysa.2013.02.031
[arXiv:1210.4588 [hep-ph]].

\bibitem{Song:2007fn}
H.~Song and U.~W.~Heinz,
Phys. Lett. B \textbf{658}, 279-283 (2008)
doi:10.1016/j.physletb.2007.11.019
[arXiv:0709.0742 [nucl-th]].

\bibitem{Shen:2011kn}
C.~Shen and U.~Heinz,
Phys. Rev. C \textbf{83}, 044909 (2011)
doi:10.1103/PhysRevC.83.044909
[arXiv:1101.3703 [nucl-th]].

\bibitem{Israel:1979wp}
W.~Israel and J.~M.~Stewart,
Annals Phys. \textbf{118}, 341-372 (1979)
doi:10.1016/0003-4916(79)90130-1

\bibitem{Bass:1998ca}
S.~A.~Bass, M.~Belkacem, M.~Bleicher, M.~Brandstetter, L.~Bravina, C.~Ernst, L.~Gerland, M.~Hofmann, S.~Hofmann and J.~Konopka, \textit{et al.}
Prog. Part. Nucl. Phys. \textbf{41}, 255-369 (1998)
doi:10.1016/S0146-6410(98)00058-1
[arXiv:nucl-th/9803035 [nucl-th]].

\bibitem{Bleicher:1999xi}
M.~Bleicher, E.~Zabrodin, C.~Spieles, S.~A.~Bass, C.~Ernst, S.~Soff, L.~Bravina, M.~Belkacem, H.~Weber and H.~Stoecker, \textit{et al.}
J. Phys. G \textbf{25}, 1859-1896 (1999)
doi:10.1088/0954-3899/25/9/308
[arXiv:hep-ph/9909407 [hep-ph]].

\bibitem{Schenke:2012wb}
B.~Schenke, P.~Tribedy and R.~Venugopalan,
Phys. Rev. Lett. \textbf{108}, 252301 (2012)
doi:10.1103/PhysRevLett.108.252301
[arXiv:1202.6646 [nucl-th]].

\bibitem{Schenke:2012hg}
B.~Schenke, P.~Tribedy and R.~Venugopalan,
Phys. Rev. C \textbf{86}, 034908 (2012)
doi:10.1103/PhysRevC.86.034908
[arXiv:1206.6805 [hep-ph]].

\bibitem{Gale:2012rq}
C.~Gale, S.~Jeon, B.~Schenke, P.~Tribedy and R.~Venugopalan,
Phys. Rev. Lett. \textbf{110}, no.1, 012302 (2013)
doi:10.1103/PhysRevLett.110.012302
[arXiv:1209.6330 [nucl-th]].

\bibitem{Monnai:2009ad}
A.~Monnai and T.~Hirano,
Phys. Rev. C \textbf{80}, 054906 (2009)
doi:10.1103/PhysRevC.80.054906
[arXiv:0903.4436 [nucl-th]].

\bibitem{Song:2009rh}
H.~Song and U.~W.~Heinz,
Phys. Rev. C \textbf{81}, 024905 (2010)
doi:10.1103/PhysRevC.81.024905
[arXiv:0909.1549 [nucl-th]].

\bibitem{Bozek:2009dw}
P.~Bozek,
Phys. Rev. C \textbf{81}, 034909 (2010)
doi:10.1103/PhysRevC.81.034909
[arXiv:0911.2397 [nucl-th]].

\bibitem{Denicol:2009am}
G.~S.~Denicol, T.~Kodama, T.~Koide and P.~Mota,
Phys. Rev. C \textbf{80}, 064901 (2009)
doi:10.1103/PhysRevC.80.064901
[arXiv:0903.3595 [hep-ph]].

\bibitem{Denicol:2010tr}
G.~S.~Denicol, T.~Kodama and T.~Koide,
J. Phys. G \textbf{37}, 094040 (2010)
doi:10.1088/0954-3899/37/9/094040
[arXiv:1002.2394 [nucl-th]].

\bibitem{Roy:2011pk}
V.~Roy and A.~K.~Chaudhuri,
Phys. Rev. C \textbf{85}, 024909 (2012)
[erratum: Phys. Rev. C \textbf{85}, 049902 (2012)]
doi:10.1103/PhysRevC.85.024909
[arXiv:1109.1630 [nucl-th]].

\bibitem{Dusling:2011fd}
K.~Dusling and T.~Sch\"afer,
Phys. Rev. C \textbf{85}, 044909 (2012)
doi:10.1103/PhysRevC.85.044909
[arXiv:1109.5181 [hep-ph]].

\bibitem{Bozek:2012fw}
P.~Bozek and W.~Broniowski,
Phys. Rev. C \textbf{85}, 044910 (2012)
doi:10.1103/PhysRevC.85.044910
[arXiv:1203.1810 [nucl-th]].

\bibitem{Noronha-Hostler:2013gga}
J.~Noronha-Hostler, G.~S.~Denicol, J.~Noronha, R.~P.~G.~Andrade and F.~Grassi,
Phys. Rev. C \textbf{88}, no.4, 044916 (2013)
doi:10.1103/PhysRevC.88.044916
[arXiv:1305.1981 [nucl-th]].

\bibitem{Rose:2014fba}
J.~B.~Rose, J.~F.~Paquet, G.~S.~Denicol, M.~Luzum, B.~Schenke, S.~Jeon and C.~Gale,
Nucl. Phys. A \textbf{931}, 926-930 (2014)
doi:10.1016/j.nuclphysa.2014.09.044
[arXiv:1408.0024 [nucl-th]].

\bibitem{Ryu:2015vwa}
S.~Ryu, J.~F.~Paquet, C.~Shen, G.~S.~Denicol, B.~Schenke, S.~Jeon and C.~Gale,
Phys. Rev. Lett. \textbf{115}, no.13, 132301 (2015)
doi:10.1103/PhysRevLett.115.132301
[arXiv:1502.01675 [nucl-th]].

\bibitem{Ryu:2017qzn}
S.~Ryu, J.~F.~Paquet, C.~Shen, G.~Denicol, B.~Schenke, S.~Jeon and C.~Gale,
Phys. Rev. C \textbf{97}, no.3, 034910 (2018)
doi:10.1103/PhysRevC.97.034910
[arXiv:1704.04216 [nucl-th]].

\bibitem{Bernhard:2019bmu}
J.~E.~Bernhard, J.~S.~Moreland and S.~A.~Bass,
Nature Phys. \textbf{15}, no.11, 1113-1117 (2019)
doi:10.1038/s41567-019-0611-8

\bibitem{JETSCAPE:2020mzn}
D.~Everett \textit{et al.} [JETSCAPE],
Phys. Rev. C \textbf{103}, no.5, 054904 (2021)
doi:10.1103/PhysRevC.103.054904
[arXiv:2011.01430 [hep-ph]].

\bibitem{JETSCAPE:2020shq}
D.~Everett \textit{et al.} [JETSCAPE],
Phys. Rev. Lett. \textbf{126}, no.24, 242301 (2021)
doi:10.1103/PhysRevLett.126.242301
[arXiv:2010.03928 [hep-ph]].

\bibitem{Nijs:2020ors}
G.~Nijs, W.~van der Schee, U.~G\"ursoy and R.~Snellings,
Phys. Rev. Lett. \textbf{126}, no.20, 202301 (2021)
doi:10.1103/PhysRevLett.126.202301
[arXiv:2010.15130 [nucl-th]].

\bibitem{Drescher:2007cd}
H.~J.~Drescher, A.~Dumitru, C.~Gombeaud and J.~Y.~Ollitrault,
Phys. Rev. C \textbf{76}, 024905 (2007)
doi:10.1103/PhysRevC.76.024905
[arXiv:0704.3553 [nucl-th]].



    
    
\bibitem{Masui:2009pw}
H.~Masui, J.~Y.~Ollitrault, R.~Snellings and A.~Tang,
Nucl. Phys. A \textbf{830}, 463C-466C (2009)
doi:10.1016/j.nuclphysa.2009.10.103
[arXiv:0908.0403 [nucl-ex]].

\bibitem{Tang:2009vp}
A.~Tang,
Nucl. Phys. A \textbf{830}, 673C-680C (2009)
doi:10.1016/j.nuclphysa.2009.10.061
[arXiv:0907.4513 [nucl-ex]].

\bibitem{Lacey:2006bc}
R.~A.~Lacey, N.~N.~Ajitanand, J.~M.~Alexander, P.~Chung, W.~G.~Holzmann, M.~Issah, A.~Taranenko, P.~Danielewicz and H.~Stoecker,
Phys. Rev. Lett. \textbf{98}, 092301 (2007)
doi:10.1103/PhysRevLett.98.092301
[arXiv:nucl-ex/0609025 [nucl-ex]].

\bibitem{Gavin:2006xd}
S.~Gavin and M.~Abdel-Aziz,
Phys. Rev. Lett. \textbf{97}, 162302 (2006)
doi:10.1103/PhysRevLett.97.162302
[arXiv:nucl-th/0606061 [nucl-th]].

\bibitem{PHENIX:2006iih}
A.~Adare \textit{et al.} [PHENIX],
Phys. Rev. Lett. \textbf{98}, 172301 (2007)
doi:10.1103/PhysRevLett.98.172301
[arXiv:nucl-ex/0611018 [nucl-ex]].

\bibitem{Kovtun:2004de}
P.~Kovtun, D.~T.~Son and A.~O.~Starinets,
Phys. Rev. Lett. \textbf{94}, 111601 (2005)
doi:10.1103/PhysRevLett.94.111601
[arXiv:hep-th/0405231 [hep-th]].

\bibitem{Bhalerao:2005mm}
R.~S.~Bhalerao, J.~P.~Blaizot, N.~Borghini and J.~Y.~Ollitrault,
Phys. Lett. B \textbf{627}, 49-54 (2005)
doi:10.1016/j.physletb.2005.08.131
[arXiv:nucl-th/0508009 [nucl-th]].


\bibitem{Teaney:2003kp}
D.~Teaney,
Phys. Rev. C \textbf{68}, 034913 (2003)
doi:10.1103/PhysRevC.68.034913
[arXiv:nucl-th/0301099 [nucl-th]].

\bibitem{Chaudhuri:2010in}
A.~K.~Chaudhuri,
Phys. Rev. C \textbf{82}, 047901 (2010)
doi:10.1103/PhysRevC.82.047901
[arXiv:1006.4478 [nucl-th]].

\bibitem{Liu:2009zzo}
J.~L.~Liu, Q.~C.~Feng, Q.~S.~Wang, G.~X.~Tang, J.~B.~Zhang and L.~Huo,
Phys. Rev. C \textbf{79}, 064905 (2009)
doi:10.1103/PhysRevC.79.064905

\bibitem{Korotkikh:2014rma}
V.~L.~Korotkikh and E.~E.~Zabrodin,
J. Phys. G \textbf{42}, no.11, 115102 (2015)
doi:10.1088/0954-3899/42/11/115102
[arXiv:1404.3050 [hep-ph]].

\bibitem{ATLAS Collaboration}
ATLAS Collaboration,
Eur.Phys.J.C 78 (2018) 12, 997
doi:10.1140/epjc/s10052-018-6468-7

\bibitem{Bhalerao:2011yg}
R.~S.~Bhalerao, M.~Luzum and J.~Y.~Ollitrault,
Phys. Rev. C \textbf{84}, 034910 (2011)
doi:10.1103/PhysRevC.84.034910
[arXiv:1104.4740 [nucl-th]].

\bibitem{Bjorken:1982qr}
J.~D.~Bjorken,
Phys. Rev. D \textbf{27}, 140-151 (1983)
doi:10.1103/PhysRevD.27.140

\bibitem{PHOBOS Collaboration:0711.3724}
PHOBOS Collaboration
Phys.Rev.C 77 (2008) 014906








\bibitem{tgmc} TGlauberMC on HepForge, version 3.0, http://www.hepforge.org/downloads/tglaubermc

\bibitem{Alver:2008aq}
B.~Alver, M.~Baker, C.~Loizides and P.~Steinberg,
[arXiv:0805.4411 [nucl-ex]].

\bibitem{Loizides:2014vua}
C.~Loizides, J.~Nagle and P.~Steinberg,
SoftwareX \textbf{1-2}, 13-18 (2015)
doi:10.1016/j.softx.2015.05.001
[arXiv:1408.2549 [nucl-ex]].

\bibitem{Kharzeev:2000ph}
D.~Kharzeev and M.~Nardi,
Phys. Lett. B \textbf{507}, 121-128 (2001)
doi:10.1016/S0370-2693(01)00457-9
[arXiv:nucl-th/0012025 [nucl-th]].

\bibitem{Moreland:2014oya}
J.~S.~Moreland, J.~E.~Bernhard and S.~A.~Bass,
Phys. Rev. C \textbf{92}, no.1, 011901 (2015)
doi:10.1103/PhysRevC.92.011901
[arXiv:1412.4708 [nucl-th]].


\bibitem{Moreland:2019szz}
J.~S.~Moreland,
[arXiv:1904.08290 [nucl-th]].

\bibitem{PHOBOS:2004vcu}
B.~B.~Back \textit{et al.} [PHOBOS],
Phys. Rev. C \textbf{72}, 051901 (2005)
doi:10.1103/PhysRevC.72.051901
[arXiv:nucl-ex/0407012 [nucl-ex]].

\bibitem{PHOBOS:2006dbo}
B.~Alver \textit{et al.} [PHOBOS],
Phys. Rev. Lett. \textbf{98}, 242302 (2007)
doi:10.1103/PhysRevLett.98.242302
[arXiv:nucl-ex/0610037 [nucl-ex]].


\bibitem{ALICE:2010suc}
K.~Aamodt \textit{et al.} [ALICE],
Phys. Rev. Lett. \textbf{105}, 252302 (2010)
doi:10.1103/PhysRevLett.105.252302
[arXiv:1011.3914 [nucl-ex]].

\bibitem{ALICE:2016ccg}
J.~Adam \textit{et al.} [ALICE],
Phys. Rev. Lett. \textbf{116}, no.13, 132302 (2016)
doi:10.1103/PhysRevLett.116.132302
[arXiv:1602.01119 [nucl-ex]].

\bibitem{ALICE:2018lao}
S.~Acharya \textit{et al.} [ALICE],
Phys. Lett. B \textbf{784}, 82-95 (2018)
doi:10.1016/j.physletb.2018.06.059
[arXiv:1805.01832 [nucl-ex]].

\bibitem{CMS:2012zex}
S.~Chatrchyan \textit{et al.} [CMS],
Phys. Rev. C \textbf{87}, no.1, 014902 (2013)
doi:10.1103/PhysRevC.87.014902
[arXiv:1204.1409 [nucl-ex]].

\bibitem{CMS:2023bvg}
A.~Tumasyan \textit{et al.} [CMS],
JHEP \textbf{2024}, no.02, 106 (2024)
doi:10.1007/JHEP02(2024)106
[arXiv:2311.11370 [nucl-ex]].

\bibitem{Nagle:2009ip}
J.~L.~Nagle, P.~Steinberg and W.~A.~Zajc,
Phys. Rev. C \textbf{81}, 024901 (2010)
doi:10.1103/PhysRevC.81.024901
[arXiv:0908.3684 [nucl-th]].





\bibitem{A. Bazavov}
A. Bazavov, Tanmoy Bhattacharya et al.
Phys. Rev. D 90, 094503 (2014)

\bibitem{Fernando G. Gardim}
F. G. Gardim, G. Giacalone, M. Luzum and Jean-Yves Ollitrault
Nature Physics 16, 615-619 (2020)
 	arXiv:1908.09728  

\bibitem{STAR Colllaboration:044902}
STAR Collaboration
Phys.Rev.C72:044902 (2005)

 \bibitem{ALICE:R. Schicker}
R. Schicker (for the ALICE Collaboration)
EPJ Web of Conferences 138, 01021 (2017)


\bibitem{ALICE:5.02}
S. Acharya et al. [ALICE Collaboration]
 Phys. Lett. B 788, 166-179 (2019)

\bibitem{Fernando G}
Fernando G. Gardim, Giuliano Giacalone. Jean-Yves Ollitrault
Phys.Lett.B 809 (2020) 135749

\bibitem{Dasgupta:2018pjm}
P.~Dasgupta, S.~De, R.~Chatterjee and D.~K.~Srivastava,
Phys. Rev. C \textbf{98}, no.2, 024911 (2018)
doi:10.1103/PhysRevC.98.024911
[arXiv:1804.02828 [nucl-th]].

\bibitem{Denicol:2014vaa}
G.~S.~Denicol, S.~Jeon and C.~Gale,
Phys. Rev. C \textbf{90}, no.2, 024912 (2014)
doi:10.1103/PhysRevC.90.024912
[arXiv:1403.0962 [nucl-th]].



\end{thebibliography}
\end{document}